\documentclass[twocolumn,aps,pre,amsmath,showpacs]{revtex4}

\usepackage{graphicx}

\def\lessgtr{\raise2.5pt\hbox{$<$}\llap{\lower2.5pt\hbox{$>$}}}
\def\gtrless{\raise2.5pt\hbox{$>$}\llap{\lower2.5pt\hbox{$<$}}}

\newcommand{\be}{\begin{equation}}
\newcommand{\ee}{\end{equation}}
\newcommand{\bea}{\begin{eqnarray}}
\newcommand{\eea}{\end{eqnarray}}

\begin{document}

\title{Structural relaxation in a system of dumbbell molecules}
\author{S.-H. Chong and W.~G{\"o}tze}
\affiliation{Physik-Department, Technische Universit{\"a}t M{\"u}nchen,
85747 Garching, Germany}
\date{\today, Phys. Rev. E, in press}

\begin{abstract}

The interaction-site-density-fluctuation correlators,
the dipole-relaxation functions, and the mean-squared
displacements of a system of symmetric dumbbells of fused
hard spheres are calculated for two representative elongations
of the molecules within the mode-coupling theory for the
evolution of glassy dynamics.
For large elongations, universal relaxation laws for
states near the glass transition are valid for parameters
and time intervals similar to the ones found for the 
hard-sphere system.
Rotation-translation coupling leads to an enlarged crossover 
interval for the mean-squared displacement of the constituent atoms between 
the end of the von Schweidler regime and the beginning of the
diffusion process.
For small elongations, the superposition principle for the 
reorientational $\alpha$-process is violated for parameters
and time intervals of interest for data analysis, and 
there is a strong breaking of the 
coupling of the $\alpha$-relaxation scale for the diffusion
process with that for representative density fluctuations
and for dipole reorientations.

\end{abstract}


\maketitle

\section{Introduction}
\label{sec:1}

Recently, the mode-coupling theory (MCT) for the evolution of glassy
dynamics in systems of spherical particles has been extended to a
theory for systems of molecules.
The fluctuations of the interaction-site densities have been used
as the basic variables to describe the structure of the system.
As a result, the known scalar MCT equations for the density fluctuations 
in simple systems have been generalized to $n$-by-$n$ matrix equations
for the interaction-site-density correlators, where $n$ denotes the number
of atoms forming the molecule.
The theory was applied to calculate the liquid-glass phase diagram and to
evaluate the glass form factors for a hard-dumbbell system 
(HDS)~\cite{Chong-MCT-dumbbell-1}.
In the following, the preceding work shall be continued by evaluating
the conventional time-dependent correlation 
functions near the liquid-glass transition.
It is the goal to examine the range of validity of the universal
relaxation laws and to identify features of the glassy dynamics
which are characteristic for molecular as opposed to atomic systems.

The dumbbells to be studied consist of two equal hard spheres of
diameter $d$ which are fused so that there is a distance
$\zeta d$, $0 \le \zeta \le 1$, between their centers.
The system's equilibrium structure for density $\rho$ is specified 
by two control parameters:
the elongation parameter $\zeta$ and the packing fraction 
$\varphi = \frac{\pi}{6} \rho d^{3} 
( 1 + \frac{3}{2} \zeta - \frac{1}{2} \zeta^{3})$.
The liquid-glass transition points $\varphi_{c}(\zeta)$ form a
non-monotonic $\varphi_{c}(\zeta)$--versus--$\zeta$ curve in the
$\varphi$--$\zeta$-control-parameter plane with a maximum for 
$\zeta$ near 0.43.
There is a second transition line $\varphi_{A}(\zeta)$ within the
glass phase $\varphi \ge \varphi_{c}(\zeta)$.
It separates a plastic-glass phase for 
$\varphi_{c}(\zeta) \le \varphi < \varphi_{A}(\zeta)$,
where dipole motion is ergodic,
from a glass for $\varphi > \varphi_{A}(\zeta)$,
where also the molecular
axes are arrested in a disordered array.
The second line $\varphi_{A}(\zeta)$ terminates 
at $\zeta = \zeta_{c} = 0.345$,
where $\varphi_{A}(\zeta_{c}) = \varphi_{c}(\zeta_{c})$ 
({\em cf}. Fig.~1 of Ref.~\onlinecite{Chong-MCT-dumbbell-1}).
In the present paper, it shall be shown that there are two
scenarios for the liquid-glass transition dynamics for 
$\zeta > \zeta_{c}$.
The first one, to be demonstrated for $\zeta = 1.0$,
deals with strong steric hindrance for reorientational motion. 
For this case, all universal relaxation laws hold within 
similar parameters and time intervals as found for the
hard-sphere system (HSS)~\cite{Franosch97,Fuchs98}.
As a new feature, there appears a very large crossover
interval for the $\alpha$-process of the 
constituent atom's mean-squared
displacement for times between the end of von Schweidler's
power law and the beginning of the diffusion regime.
The other scenario, to be shown for $\zeta = 0.4$, 
deals with weak steric hindrance for reorientational motion.
The universal laws for the reorientational $\alpha$-process are
restricted to such narrow $\varphi - \varphi_{c}(\zeta)$ intervals
that they are practically irrelevant for the interpretation of 
data obtained by molecular-dynamics simulations or by
presently used spectrometers.

The paper is organized as follows.
In Sec.~\ref{subsec:2A}, the MCT equations of motion
for the coherent and incoherent density correlation functions for
the HDS are listed.
Section~\ref{subsec:2B} contains known universal laws for the MCT glass transition.
The next section presents the new results for representative
density-fluctuation correlators (Sec.~\ref{subsec:3A}), for the
dipole dynamics (Sec.~\ref{subsec:3B}),
for the mean-squared displacements (Sec.~\ref{subsec:3C}),
and for the $\alpha$-relaxation scales (Sec.~\ref{subsec:3D}).
The findings are summarized in Sec.~\ref{sec:4}.

\section{MCT equations for the HDS}
\label{sec:2}

\subsection{Equations of motion for the density correlators}
\label{subsec:2A}

In this subsection, the basic equations for the system of symmetric
hard dumbbells are noted.
They have been derived in Ref.~\onlinecite{Chong-MCT-dumbbell-1},
and their solutions underlie all results to be
discussed in the present paper.

If ${\vec r}_{i}^{\, a}$, $a = A$ or $B$,
denote the interaction-site centers of the $i$th $AB$ 
dumbbell molecule,
the interaction-site-density fluctuations for wave vector ${\vec q}$
read
$\rho_{\vec q}^{a} = \sum_{i} \exp( i {\vec q} \cdot {\vec r}_{i}^{\, a} )$.
Similarly, the tagged-molecule-density fluctuations read
$\rho_{{\vec q},s}^{a} = \exp( i {\vec q} \cdot {\vec r}_{s}^{\, a})$
with ${\vec r}_{s}^{\, a}$ denoting the interaction-site positions
of the tagged molecule.
It is convenient to transform to total number densities
$\rho_{\vec q}^{N} = (\rho_{\vec q}^{A} + \rho_{\vec q}^{B})/\sqrt{2}$
and ``charge'' densities
$\rho_{\vec q}^{Z} = (\rho_{\vec q}^{A} - \rho_{\vec q}^{B})/\sqrt{2}$.
Similar definitions are used for the tagged-molecule densities.
The top-down symmetry of the molecules implies
vanishing cross correlations
$\langle \rho_{\vec q}^{N}(t)^{*} \rho_{\vec q}^{Z}(0) \rangle =
\langle \rho_{{\vec q},s}^{N}(t)^{*} \rho_{{\vec q},s}^{Z}(0) \rangle =0$
and also the identity
$\langle \rho_{\vec q}^{Z}(t)^{*} \rho_{\vec q}^{Z}(0) \rangle /
\langle | \, \rho_{\vec q}^{Z} \,|^{2} \rangle =
\langle \rho_{{\vec q},s}^{Z}(t)^{*} \rho_{{\vec q},s}^{Z}(0) \rangle /
\langle | \, \rho_{{\vec q},s}^{Z} \,|^{2} \rangle$.
There is only one independent coherent density correlator,
which shall be used as normalized function
\be
\phi_{q}^{N}(t) = 
\langle \rho_{\vec q}^{N}(t)^{*} \rho_{\vec q}^{N}(0) \rangle /
\langle | \, \rho_{\vec q}^{N} \,|^{2} \rangle.
\label{eq:phi-N-def}
\ee
There are two incoherent density correlators
\be
\phi_{q,s}^{x}(t) = 
\langle \rho_{{\vec q},s}^{x}(t)^{*} \rho_{{\vec q},s}^{x}(0) \rangle /
\langle | \, \rho_{{\vec q},s}^{x} \,|^{2} \rangle, \quad
x = N, Z. 
\ee
Because of rotational invariance, the density correlators
depend on the wave-vector modulus
$q = |\, {\vec q} \,|$ only.
The short-time asymptotics of the correlators read:
$\phi_{q}^{N}(t) = 1 - \frac{1}{2} (\Omega_{q}^{N} t)^{2} + O(|t|^{3})$ and
$\phi_{q,s}^{x}(t) = 1 - \frac{1}{2} (\Omega_{q,s}^{x} t)^{2} + O(|t|^{3})$.
The characteristic frequencies are given by the formulas:
$(\Omega_{q}^{N})^{2} = q^{2} \{ 
v_{T}^{2} [ 1 + j_{0}(q \zeta d)] +
(v_{R}^{2} \zeta^{2} d^{2} / 6) [1 - j_{0}(q \zeta d) - j_{2}(q \zeta d)]
\} / S_{q}^{N}$ and
$(\Omega_{q,s}^{x})^{2} = q^{2} \{ 
v_{T}^{2} [ 1 \pm j_{0}(q \zeta d)] +
(v_{R}^{2} \zeta^{2} d^{2} / 6) [1 \mp j_{0}(q \zeta d) \mp j_{2}(q \zeta d)]
\} / w_{q}^{x}$ 
for $x = N, Z$.
Here $j_{\ell}$ denotes the spherical Bessel function for index $\ell$,
$S_{q}^{N}$ abbreviates the total-density structure factor,
and $w_{q}^{x} = 1 \pm j_{0}(q \zeta d)$ for $x = N, Z$
are the intramolecular structure factors.
Furthermore, $v_{T} = \sqrt{k_{B}T/2m}$ denotes the thermal velocity
for translation of the molecule of atomic masses $m$ at 
temperature $T$, and 
$v_{R} = (2 / \zeta d) v_{T}$ is the thermal velocity for rotations.
The Zwanzig-Mori equations of motion~\cite{Hansen86}
for the specified correlation functions are:
\bea
& &
\partial_{t}^{2} \phi^{N}_{q}(t) + (\Omega_{q}^{N})^{2} \, \phi^{N}_{q}(t) 
\nonumber \\
& & \quad \,\,
+ \,
(\Omega_{q}^{N})^{2} 
\int_{0}^{t} dt' \, m^{N}_{q}(t-t') \, \partial_{t'} \phi^{N}_{q}(t') = 0,
\label{eq:GLE-N}
\\
& &
\partial_{t}^{2} \phi^{x}_{q,s}(t) + 
(\Omega^{x}_{q,s})^{2} \, \phi^{x}_{q,s}(t) 
\nonumber \\
& & \quad \,\,
+ \,
(\Omega^{x}_{q,s})^{2} 
\int_{0}^{t} dt' \, m^{x}_{q,s}(t-t') \, 
\partial_{t'} \phi^{x}_{q,s}(t') = 0,
\label{eq:GLE-sNZ}
\eea
for $x = N, Z$. 
All complications of the dynamics of the many-particle problem
are hidden in the relaxation kernels $m_{q}^{N}(t)$ and
$m_{q,s}^{x}(t)$.

Within MCT, the relaxation kernel $m_{q}^{N}(t)$ is expressed
as functional of the density correlators:
\begin{subequations}
\label{eq:MCT-N}
\be
m_{q}^{N}(t) = {\cal F}_{q}^{N}[\phi^{N}(t)].
\label{eq:MCT-N-a}
\ee
The mode-coupling functional reads
${\cal F}_{q}^{N}[\tilde{f}] = 
\frac{1}{2} \int d{\vec k} \, V^{N}({\vec q}; {\vec k}, {\vec p} \,) 
\tilde{f}_{k} \tilde{f}_{p}$
where 
${\vec k} + {\vec p} = {\vec q}$, and
the positive coupling vertices are given by the density $\rho$,
the structure factor $S_{q}^{N}$, and the direct
correlation function $c_{q}^{N}$
({\em cf}. Eq.~(20b) of Ref.~\onlinecite{Chong-MCT-dumbbell-1}).
Wave-vector integrals are converted into discrete sums by 
introducing some upper cutoff $q^{*}$ and using
a grid of $M$ equally spaced values for the modulus:
$qd = h/2, 3h/2, \cdots, (q^{*}d - h/2)$.
Thus, $q$ can be considered as a label for an array of $M$ values.
Equation~(\ref{eq:GLE-N}) then represents a set of $M$ equations which are 
coupled by the polynomial
\be
{\cal F}_{q}^{N}[\tilde{f}] =
\sum_{k,p} V_{q, kp} \, \tilde{f}_{k} \, \tilde{f}_{p}.
\label{eq:MCT-N-b}
\ee
\end{subequations}
It is an elementary task to express the $M^{3}$ positive 
coefficients $V_{q,kp}$ in terms of the
$V^{N}({\vec q}; {\vec k}, {\vec p} \,)$
({\em cf}. Eq.~(7) of Ref.~\onlinecite{Franosch97}).
Similarly, one derives 
functionals for the kernels for the tagged-molecule functions:
\begin{subequations}
\label{eq:MCT-sNZ}
\be
m_{q,s}^{x} = {\cal F}_{q,s}^{x}[ \phi_{s}^{x}(t), \phi^{N}(t) ], \quad
x = N, Z. 
\label{eq:MCT-sNZ-a}
\ee
Again, the functionals ${\cal F}_{q,s}^{x}$ are given in terms of
the equilibrium structure functions
({\em cf}. Eq.~(23) of Ref.~\onlinecite{Chong-MCT-dumbbell-1}).
After the discretization, the mode-coupling polynomial has the form
\be
{\cal F}_{q,s}^{x}[\tilde{f}_{s}^{x}, \tilde{f}] =
\sum_{k,p} V_{q,kp}^{x} \, \tilde{f}_{k,s}^{x} \, \tilde{f}_{p}, \quad
x = N, Z. 
\label{eq:MCT-sNZ-b}
\ee
\end{subequations}

The following work is done for a cutoff $q^{*}d = 40$
and $M = 100$ wave-vector moduli.
The structure factors are evaluated within the 
reference-interaction-site-model (RISM) 
theory~\cite{Hansen86,Chandler72,Lowden73}.
In Ref.~\onlinecite{Chong-MCT-dumbbell-1}, the details of the
static correlation functions have been discussed.
Equations~(\ref{eq:GLE-N}) and (\ref{eq:MCT-N}) are closed
non-linear integro-differential equations to calculate the
$M$ correlators $\phi_{q}^{N}(t)$.
Using these correlators as input,
Eqs.~(\ref{eq:GLE-sNZ}) and (\ref{eq:MCT-sNZ}) are closed
equations to evaluate the $M$ correlators $\phi_{q,s}^{x}(t)$
for $x = N$ and $Z$.
The mathematical structure of 
Eqs.~(\ref{eq:GLE-N}) and (\ref{eq:MCT-N}) is identical to that
discussed earlier for the density correlators of the
HSS~\cite{Franosch97}.
The differences are merely the values of the frequencies
$\Omega_{q}^{N}$ and the values for the coupling constants
$V_{q,kp}$.
A similar statement holds for 
Eqs.~(\ref{eq:GLE-sNZ}) and (\ref{eq:MCT-sNZ}),
which are the analogues of the equations for the motion of a sphere
in the HSS~\cite{Fuchs98}.
Therefore, all general results discussed 
previously~\cite{Franosch97,Fuchs98} hold for the
present theory as well.

In the rest of the paper, the diameter of the constituent atoms
shall be chosen
as unit of length, $d = 1$, and the unit of time is chosen so,
that $v_{T} = 1$.

\subsection{Universal laws}
\label{subsec:2B}

This subsection compiles the universal laws for the dynamics near the
liquid-glass transition.
They will be used in the following Sec.~\ref{sec:3} to analyze the 
numerical solutions of the equations of motion.
The derivation of these laws is discussed comprehensively
in Refs.~\onlinecite{Franosch97} and \onlinecite{Fuchs98},
where also the earlier literature on this subject is cited.

From the mode-coupling functional in Eq.~(\ref{eq:MCT-N-b}),
one calculates an $M$-by-$M$ matrix
$C_{qk} = 
\{ \partial {\cal F}_{q}^{N}[f^{N}] / \partial f_{k}^{N} \} (1-f_{k}^{N})^{2}$.
Here $f_{k}^{N} = \phi_{k}^{N}(t \to \infty)$ denote
the nonergodicity parameters for the density fluctuations
of the glass states.
This matrix has a non-degenerate maximum eigenvalue $E \le 1$.
The transition is characterized by $E^{c} = 1$.
Here and in the following, the superscript $c$ indicates that
the quantity is evaluated for $\varphi = \varphi_{c}(\zeta)$.
Let $e$ and $\hat{e}$ denote the right and left eigenvectors,
respectively, at the critical points:
$\sum_{k} C_{qk}^{c} e_{k} = e_{q}$, 
$\sum_{q} \hat{e}_{q} C_{qk}^{c} = \hat{e}_{k}$.
The eigenvectors are fixed uniquely by requiring 
$e_{q} > 0$, $\hat{e}_{q} > 0$,
$\sum_{q} \hat{e}_{q} e_{q} = 1$, and
$\sum_{q} \hat{e}_{q} (1-f_{q}^{Nc}) e_{q} e_{q} = 1$.
These eigenvectors are obtained as a byproduct of the numerical 
determination of $\varphi_{c}(\zeta)$ described in 
Ref.~\onlinecite{Chong-MCT-dumbbell-1}.
They are used to evaluate the critical amplitude
\be
h_{q}^{N} = (1 -f_{q}^{Nc})^{2} \, e_{q},
\ee
and the exponent parameter $\lambda$, $1/2 \le \lambda < 1$,
\be
\lambda =
\frac{1}{2} 
\sum_{q,k,p} \hat{e}_{q} \,
\{ \partial^{2} {\cal F}_{q}^{Nc}[f^{Nc}] \, / \,
\partial f_{k}^{Nc} \, \partial f_{p}^{Nc} \} \,
h_{k}^{N} \, h_{p}^{N}.
\label{eq:lambda}
\ee
Furthermore, a smooth function $\sigma$ of the control parameters
$\zeta$ and $\varphi$ is defined by
\begin{subequations}
\label{eq:sigma}
\be
\sigma = \sum_{q} \hat{e}_{q} \,
\{ {\cal F}_{q}^{N}[f^{Nc}] - {\cal F}_{q}^{Nc}[f^{Nc}] \}.
\label{eq:sigma-a}
\ee
In a leading expansion for small $\varphi - \varphi_{c}(\zeta)$, one can write
\be
\sigma = C \, \epsilon, \quad
\epsilon = (\varphi - \varphi_{c}(\zeta)) \, / \, \varphi_{c}(\zeta).
\label{eq:sigma-b}
\ee
\end{subequations}
Glass states are characterized by $\sigma > 0$,
liquid states by $\sigma < 0$, and $\sigma = 0$ specifies the
transition.
The nonergodicity parameter exhibits a square-root singularity.
In a leading-order expansion for small $\epsilon$, one gets
\be
f_{q}^{N} = f_{q}^{Nc} + h_{q}^{N} \, \sqrt{\sigma / (1 - \lambda)}, \quad
\sigma > 0, \quad
\sigma \to 0.
\label{eq:fq}
\ee
At the transition, the correlator exhibits a 
power-law decay which is specified by the
critical exponent $a$, $0 < a < 1/2$.
In a leading-order expansion in $(1/t)^{a}$, one gets
\be
\phi_{q}^{N}(t) = f_{q}^{Nc} + h_{q}^{N} \, (t_{0}/t)^{a}, \quad
\sigma = 0, \quad
(t/t_{0}) \to \infty.
\label{eq:phi-N-critical}
\ee
Here, $t_{0}$ is a time scale for the transient dynamics.
The exponent $a$ is determined from the equation
$\Gamma(1-a)^{2} / \Gamma(1-2a) = \lambda$,
where $\Gamma$ denotes the gamma function.

Let us consider the correlator 
$\phi_{Y}(t) = \langle Y(t)^{*} Y(0) \rangle /
\langle | Y |^{2} \rangle$
of some variable $Y$ coupling to the density fluctuations.
Its nonergodicity parameter $f_{Y} = \phi_{Y}(t \to \infty)$
obeys an equation analogous to Eq.~(\ref{eq:fq}):
$f_{Y} = f_{Y}^{c} + h_{Y} \sqrt{\sigma/(1 - \lambda)} + O(\sigma)$.
The critical nonergodicity parameter $f_{Y}^{c} > 0$ and the
critical amplitude $h_{Y} > 0$ are equilibrium quantities
to be calculated from the relevant mode-coupling functionals
at the critical point $\varphi = \varphi_{c}(\zeta)$.
If $Y = \rho_{{\vec q},s}^{x}$, the correlator $\phi_{Y}(t)$
refers to the tagged-molecule-density fluctuations,
$\phi_{q,s}^{x}(t)$.
Their nonergodicity parameters $f_{Y} = f_{q,s}^{x}$ have been 
discussed in Ref.~\onlinecite{Chong-MCT-dumbbell-1},
and the explicit formulas for the evaluation of 
$h_{Y} = h_{q,s}^{x}$ can be inferred from Ref.~\onlinecite{Fuchs98}.
For small values of $\epsilon$,
there is a large time interval, where 
$\phi_{Y}(t)$ is close to $f_{Y}^{c}$.
Solving the equations of motion asymptotically 
for this plateau regime, one gets in leading order in the
small quantities
$\phi_{Y}(t) - f_{Y}^{c}$
the factorization theorem: 
\be
\phi_{Y}(t) - f_{Y}^{c} = h_{Y} \, G(t).
\label{eq:first-a}
\ee
The function $G(t)$ is the same for all variables $Y$.
It describes the complete dependence on time and 
on control parameters via the first scaling law:
\be
G(t) = \sqrt{| \, \sigma \, |} \, g_{\pm}(t/t_{\sigma}), \quad
\sigma \gtrless 0.
\label{eq:first-b}
\ee
Here
\be
t_{\sigma} = t_{0} \, / \, | \, \sigma \, |^{\delta}, \quad
\delta = 1/2a,
\label{eq:first-c}
\ee
is the first critical time scale.
The master functions $g_{\pm}(\hat{t})$ are determined by $\lambda$;
they can easily be evaluated numerically~\cite{Goetze90}.
One gets $g_{\pm}(\hat{t} \to 0) = 1 / \hat{t}^{a}$, so that
Eq.~(\ref{eq:phi-N-critical}) for $Y = \rho_{\vec q}^{N}$
is reproduced for fixed large $t$
and $\sigma$ tending to zero.
Since $g_{+}(\hat{t} \to \infty) = 1 / \sqrt{1-\lambda}$, 
also Eq.~(\ref{eq:fq}) is reproduced. 

One finds for the large rescaled time $\hat{t}$:
$g_{-}(\hat{t} \to \infty) = - B \hat{t}^{b} + O(1/\hat{t}^{b})$.
The anomalous exponent $b$, $0 < b \le 1$, which is called 
the von Schweidler exponent, is to be calculated from $\lambda$ 
via the equation
$\Gamma(1+b)^{2} / \Gamma(1+2b) = \lambda$.
The constant $B$ is of order unity, and is also fixed by 
$\lambda$~\cite{Goetze90}. 
Substituting this result into 
Eqs.~(\ref{eq:first-a}) and (\ref{eq:first-b}), one obtains
von Schweidler's law for the decay of the liquid correlator below the
plateau $f_{Y}^{c}$:
\be
\phi_{Y}(t) = f_{Y}^{c} - h_{Y} \, (t/t_{\sigma}^{\prime})^{b}, \quad
t_{\sigma} \ll t, \quad
\sigma \to -0.
\label{eq:phi-Y-von}
\ee
The control-parameter dependence is given via
the second critical time scale $t_{\sigma}^{\prime}$: 
\be
t_{\sigma}^{\prime} = t_{0} \, B^{-1/b} \, / \, |\sigma|^{\gamma}, \quad
\gamma = (1/2a) + (1/2b).
\label{eq:t-sigma-prime}
\ee
The leading corrections to the preceding formulas
for fixed $\hat{t} = t / t_{\sigma}$ are
proportional to $|\, \sigma \,|$.
They are specified by two correction amplitudes and additional scaling 
functions $h_{\pm}(t/t_{\sigma})$~\cite{Franosch97}.
The dynamical process described by the cited results is called
$\beta$-process.
The $\beta$-correlator $G(t)$ describes in leading order the decay
of the correlator towards the plateau value $f_{Y}^{c}$ within
the interval $t_{0} \ll t \ll t_{\sigma}$.
The glass correlators arrest at $f_{Y}$ for $t \gg t_{\sigma}$.
In the limit $\sigma \to -0$, all correlators cross
their plateau $f_{Y}^{c}$ at the same time $\tau_{\beta}$, given by
\be
\tau_{\beta} = \hat{t}_{-} \, t_{\sigma}.
\label{eq:tau-beta}
\ee
Here, the number $\hat{t}_{-}$ is fixed by $\lambda$ via
$g_{-}(\hat{t}_{-}) = 0$.

The decay of $\phi_{Y}(t)$ below the plateau $f_{Y}^{c}$ is called the 
$\alpha$-process for variable $Y$. 
For this process, there holds the second scaling law 
in leading order for $\sigma \to -0$:
\be
\phi_{Y}(t) = \tilde{\phi}_{Y}(\tilde{t}), \quad
\tilde{t} = t / t_{\sigma}^{\prime}, \quad
t_{\sigma} \ll t,
\label{eq:phi-Y-second}
\ee
which is also referred to as the superposition principle. 
The control-parameter-independent shape function 
$\tilde{\phi}_{Y}(\tilde{t})$ is to be evaluated from the mode-coupling
functionals at the critical point. 
For short rescaled times $\tilde{t}$,
one gets 
$\tilde{\phi}_{Y}(\tilde{t}) = f_{Y}^{c} - h_{Y} \tilde{t}^{b} + O(\tilde{t}^{2b})$, 
so that Eq.~(\ref{eq:phi-Y-von}) is reproduced.
The ranges of applicability of the first and the second scaling laws
overlap;
both scaling laws yield von Schweidler's law for 
$t_{\sigma} \ll t \ll t_{\sigma}^{\prime}$.

The superposition principle implies
coupling of the $\alpha$-relaxation time scales or 
relaxation rates of all the variables
in the following sense. 
Let us characterize the long-time decay of $\phi_{Y}(t)$ 
in the liquid by some time $\tau_{\alpha}^{Y}$.
For example, 
as is occasionally done in analyzing molecular-dynamics simulation
data, it may be defined as the center of the $\alpha$
process, i.e. as the time at which the correlator
has decayed to 50\% of its plateau value:
\be
\phi_{Y}(\tau_{\alpha}^{Y}) = f_{Y}^{c} / 2.
\label{eq:tau-alpha}
\ee
This time diverges upon approaching the glass transition:
in the leading asymptotic limit for $\sigma \to -0$,
one finds $\tau_{\alpha}^{Y} = C_{Y} t_{\sigma}^{\prime}$.
Here, the constant $C_{Y}$ is defined by
$\tilde{\phi}_{Y}(C_{Y}) = f_{Y}^{c}/2$. 
All times or rates are proportional to each other, and 
follow a power law specified by the exponent $\gamma$
defined in Eq.~(\ref{eq:t-sigma-prime}):
\be
1/\tau_{\alpha}^{Y} = \Gamma_{Y} \, | \, \epsilon \, |^{\gamma}.
\label{eq:scale-coupling}
\ee
The constants of proportionality $C_{Y}$ or $\Gamma_{Y}$
depend on the variable $Y$ and on the precise convention
for the definition of $\tau_{\alpha}^{Y}$, such as
Eq.~(\ref{eq:tau-alpha}).
All the time scales or rates are coupled in the sense that
the ratio $C_{Y}/C_{Y'}$ or $\Gamma_{Y}/\Gamma_{Y'}$ 
for two different variables $Y$ and $Y'$
becomes independent of the control parameters in the limit
$\sigma \to -0$. 

\begin{figure}
\includegraphics[width=0.8\linewidth]{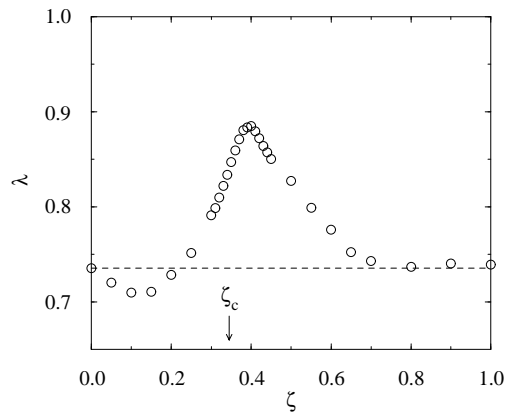}
\caption{
Exponent parameter $\lambda$ as function of the elongation $\zeta$.
The arrow indicates the critical elongation $\zeta_{c} = 0.345$
separating the two glass phases.
The dashed horizontal line marks the value $\lambda = 0.736$
for the hard-sphere system.}
\label{fig:lambda}
\end{figure}

Figure~\ref{fig:lambda} shows that for large elongations
of the molecules, say $\zeta > 0.6$,
and for very small elongations, say $\zeta < 0.2$,
the exponent parameter $\lambda$ is close to 
the value $0.736 \pm 0.003$ for the HSS ($\zeta = 0$).
For $\zeta \sim \zeta_{c}$, 
the two contributions to the structure-factor peak for 
angular-momentum indices $\ell = 0$ and $\ell = 2$
are of equal importance~\cite{Chong-MCT-dumbbell-1}.
For such elongations, $\lambda$ increases considerably.
Unfortunately, Eq.~(\ref{eq:lambda}) is so involved that
we cannot trace back the increase of $\lambda$ to the
underlying variations of the structure factor. 

The accurate determination of the time scale $t_{0}$
is cumbersome.
It is given by the constant plateau value of the function
$\Psi(t) = \{ [\phi_{q}^{Nc}(t) - f_{q}^{Nc}] / h_{q}^{N} \}^{1/a} \, t$
within the time interval where the leading-order 
Eq.~(\ref{eq:phi-N-critical}) is valid.
The leading corrections to this law vary proportional to 
$(t_{0}/t)^{2a}$ and cause deviations from the plateau at small times.
The unavoidable errors in the determination of the critical value
$\varphi_{c}(\zeta)$ cause $\sigma \ne 0$ in 
Eq.~(\ref{eq:sigma-b}). 
Thus, $\Psi(t)$ increases proportional to $t$ or $-t$
for $t \to \infty$ if $\sigma > 0$ or $\sigma < 0$, respectively.
For example, 
to determine $t_{0}$ for $\zeta = 1.0$
with an error estimated to be smaller than 3\%,
we have fixed $\varphi_{c}(\zeta)$ with 9 relevant digits.
The plateau regime of $\Psi(t)$ is largest for $q = 9.8$,
where $f_{q}^{Nc}$ has an intermediate value;
it extends from $t \sim 10^{4}$ to $t \sim 10^{9}$.
We have checked that results for $q = 3.0$, $7.4$, $13.0$, and $16.2$
lead to the same $t_{0}$ within the specified error. 
A similar statement holds for the 
determination of $t_{0}$ for other elongations. 

Table~\ref{table:1} compiles the parameters characterizing the
universal formulas for the three elongations to be considered.

\begin{table}
\caption{
Parameters characterizing the
MCT-liquid-glass-transition dynamics
for systems with elongations
$\zeta = 0.0$, $0.4$, and $1.0$.}
\begin{ruledtabular}
\begin{tabular}{cccccccccc}
$\zeta$ & $\varphi_{c}$ & $C$  & $t_{0}$ & $\lambda$ & $a$   & $b$   & $\gamma$ & $B$   & $\hat{t}_{-}$ \\
\colrule
0.0     & 0.530         & 1.54 & 0.0220  & 0.736     & 0.311 & 0.582 & 2.46     & 0.838 & 0.703         \\
0.4     & 0.675         & 1.81 & 0.0123  & 0.885     & 0.222 & 0.330 & 3.77     & 2.54  & 0.110         \\
1.0     & 0.565         & 1.90 & 0.0139  & 0.739     & 0.310 & 0.576 & 2.48     & 0.857 & 0.687         \\
\end{tabular}
\end{ruledtabular}
\label{table:1}
\end{table}

\section{Results for the structural relaxation}
\label{sec:3}

\subsection{Density correlators}
\label{subsec:3A}

\begin{figure}
\includegraphics[width=0.8\linewidth]{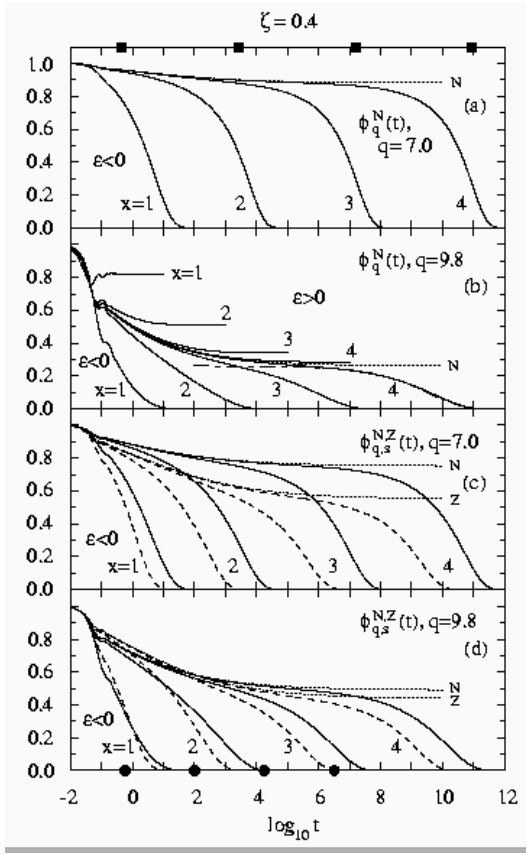}
\caption{Coherent density correlators $\phi_{q}^{N}(t)$
[solid lines in (a) and (b)],
tagged-molecule density correlators
$\phi_{q,s}^{N}(t)$ [solid lines in (c) and (d)],
and tagged-molecule charge-density correlators
$\phi_{q,s}^{Z}(t)$ [dashed lines in (c) and (d)]
for the elongation $\zeta = 0.4$ as function of $\log_{10}t$ for
two intermediate wave numbers $q = 7.0$ and 9.8.
The decay curves at the
critical packing fraction $\varphi_{c}$ are shown as dotted lines
and marked by $N$ and $Z$ for the total density and charge
density correlators, respectively.
Glass curves ($\epsilon>0$) are shown only for $\phi_{q}^{N}(t)$
with $q = 9.8$ in order to avoid overcrowding of the
figure.
The packing fractions are parameterized by
$\epsilon = (\varphi - \varphi_c) / \varphi_c = \pm 10^{-x}$,
and $x = 1$, 2, 3, 4 are chosen.
The dashed-dotted line in (b) is a Kohlrausch-law fit
for the $x = 4$ $\alpha$-process:
$f_{q}^{Nc} \exp[ - (t/\tau)^{\beta} ]$ with
$\beta = 0.40$.
The filled circles and squares mark the characteristic
times $t_{\sigma}$ and $t'_{\sigma}$, respectively,
defined in Eqs.~(\protect\ref{eq:first-c}) and
(\protect\ref{eq:t-sigma-prime})
for $x = 1$, 2, 3 and 4.
Here and in the following figures,
the diameter of the constituent atoms is chosen as the unit of
length, $d = 1$, and the unit of time is chosen so that
$v_{T} = 1$.}
\label{fig:NZ-t-weak}
\end{figure}

\begin{figure}
\includegraphics[width=0.8\linewidth]{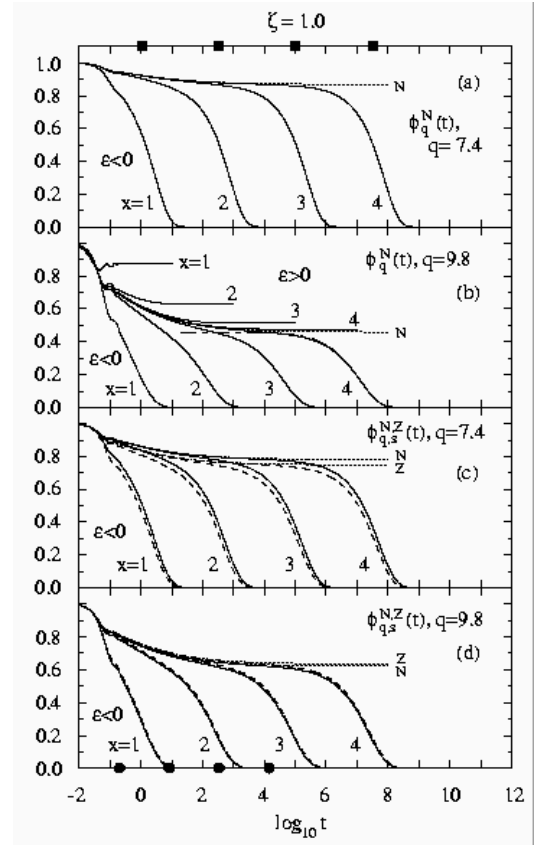}
\caption{
Results as in Fig.~\protect\ref{fig:NZ-t-weak},
but for the elongation $\zeta = 1.0$ for wave numbers
$q = 7.4$ and $9.8$, and a Kohlrausch exponent
$\beta = 0.68$.}
\label{fig:NZ-t-strong}
\end{figure}

Figures~\ref{fig:NZ-t-weak} and \ref{fig:NZ-t-strong} 
demonstrate the coherent density correlators $\phi_{q}^{N}(t)$ 
and the tagged-molecule correlation functions
$\phi_{q,s}^{N}(t)$ and $\phi_{q,s}^{Z}(t)$
near the liquid-glass transition for the
elongations $\zeta = 0.4$ and $1.0$, respectively. 
The results are for two representative wave numbers $q$: 
the wave number $q \approx 7.0$ is close to the
first peak, and $q \approx 9.8$
is near the first minimum of $S_{q}^{N}$
({\em cf}. Fig.~2 of Ref.~\onlinecite{Chong-MCT-dumbbell-1}).
The oscillatory transient dynamics occurs within the
short-time window $t < 1$. 
The control-parameter sensitive
glassy dynamics occurs for longer times for packing
fractions $\varphi$ near $\varphi_c$. 
At the transition point $\varphi = \varphi_c$, the correlators decrease
in a stretched manner towards the plateau values 
($f_{q}^{Nc}$, $f_{q,s}^{Nc}$ or $f_{q,s}^{Zc}$)
as shown by the dotted lines. 
Increasing $\varphi$
above $\varphi_c$, the long-time limits increase, as shown for
the coherent correlators $\phi_{q}^{N}(t)$ for $q = 9.8$. 
Decreasing $\varphi$ below $\varphi_c$, 
the correlators cross their plateaus at
some time $\tau_\beta$, and then decay towards zero.
For small $|\, \varphi - \varphi_{c} \,|$, 
$\tau_{\beta}$ is given by 
Eq.~(\ref{eq:tau-beta}) for all the correlators.
The decay from the plateau to zero is called the $\alpha$-process.
It is characterized, e.g., by the $\alpha$-time scale $\tau_{\alpha}$
defined as in Eq.~(\ref{eq:tau-alpha}) for each correlator.
Thus, upon decreasing $\varphi_{c} - \varphi$ towards zero, 
the time scales $\tau_{\beta}$ and
$\tau_{\alpha}$ increase towards infinity proportional to $t_\sigma$
and $t_\sigma^\prime$, respectively.
The two-step-relaxation scenario emerges, 
because the ratio of the scales
$\tau_{\alpha}/\tau_{\beta}$ increases as well.
Figures~\ref{fig:NZ-t-weak} and \ref{fig:NZ-t-strong}
exemplify the standard MCT-bifurcation scenario. 
For small $|\, \varphi - \varphi_c \,|$, 
the results can be described in terms of scaling laws
mentioned in Sec.~\ref{subsec:2B}. 
This was demonstrated comprehensively in
Refs.~\onlinecite{Franosch97} and \onlinecite{Fuchs98} for the HSS, 
and the discussion shall not be repeated here. 

For $\zeta = 1.0$ and $q \geq 5$, 
the plateaus for the tagged molecule's 
total-density and charge-density fluctuations are very close 
to each other: $f_{q,s}^{Nc} \approx f_{q,s}^{Zc}$
({\em cf}. Figs.~12 and 13 of Ref.~\onlinecite{Chong-MCT-dumbbell-1}).
Figures~\ref{fig:NZ-t-strong}(c) and \ref{fig:NZ-t-strong}(d)
demonstrate that also the dynamics is nearly the same, 
$\phi_{q,s}^{N}(t) \approx \phi_{q,s}^{Z}(t)$. 
This means that
for $q \zeta \ge 5$ and for strong steric hindrance, 
the cross correlations $F_{q,s}^{AB}(t)$ are very small. 
The reason is that the
intramolecular correlation factors $j_\ell (q \zeta /2)$ are
small, and thus interference effects between the density
fluctuations of the two interaction sites are suppressed.
Coherence effects can be expected only for small wave numbers.
For this case, 
the functions can be understood in terms of
their small-$q$ asymptotes.
The latter are determined by the
dipole correlator and the mean-squared 
displacements~\cite{Chong01}, and their
results shall be discussed in the following two subsections. 

Figures~\ref{fig:NZ-t-weak}(c) and \ref{fig:NZ-t-weak}(d) deal 
with tagged molecule's correlators of 
weak steric hindrance.
In this case, the charge-density fluctuations 
behave quite differently from the total density
fluctuations. 
The most important origin of this difference is the
reduction of the mode-coupling vertex in 
Eq.~(\ref{eq:MCT-sNZ-b})
for $x = Z$ relative to the one for $x = N$. 
For small elongations, 
the effective potentials from the surroundings
for the tagged molecule's reorientation are small.
Therefore, $f_{q,s}^{Zc}$ decreases strongly relative to
$f_{q,s}^{Nc}$ for $\zeta$ decreasing towards $\zeta_c$
({\em cf}. Fig.~13 of Ref.~\onlinecite{Chong-MCT-dumbbell-1}).
For $\zeta < \zeta_{c}$, the charge-density fluctuations
relax to zero as in a normal liquid.
This implies, as precursor phenomenon, that
the $\alpha$-time scale $\tau_{\alpha}^{sZ}$ of the
charge-density fluctuations 
shortens relative to the scale $\tau_{\alpha}^{sN}$ for the
total density fluctuations. 
Thus, the differences between $\phi_{q,s}^{N}(t)$ and $\phi_{q,s}^{Z}(t)$
for small $\zeta$ shown in 
Figs.~\ref{fig:NZ-t-weak}(c) and \ref{fig:NZ-t-weak}(d) 
are due to disturbances of the 
standard transition scenario by the nearby 
type-$A$ transition from a normal glass to a plastic glass. 

The stretching of the relaxation process is much more pronounced
for the $\zeta = 0.4$ system than for the $\zeta = 1.0$ system.
The wave vectors $q = 7.0$ and 7.4 refer to the structure-factor
peak position for $\zeta = 0.4$ and 1.0, respectively,
and the corresponding plateau values $f_{q}^{Nc}$ are almost
the same.
Figure~\ref{fig:NZ-t-weak}(a) 
demonstrates that the $\zeta = 0.4$ correlator for 
$\epsilon = - 10^{-4}$ requires a time increase by 2.3 decades
for the decay from 90\% to 10\% of the plateau value $f_{q}^{Nc}$.
The corresponding decay interval for the $\zeta = 1.0$ correlator
is 1.7 decades, as shown in Fig.~\ref{fig:NZ-t-strong}(a).
Thus, the specified time interval for the $\alpha$-process is 4 times
larger for the small than for the large elongation.
Often, stretching is quantified by the exponent $\beta_{Y}$ of the
Kohlrausch-law fit for the $\alpha$-process:
$\phi_{Y}(t \ge \tau_{\beta}) \propto
\exp[-(t/\tau_{\alpha}^{\prime})^{\beta_{Y}}]$.
Such fits are shown by the dashed-dotted lines for $\phi_{q}^{N}(t)$
for $q = 9.8$ and $\epsilon = -10^{-4}$ in 
Figs.~\ref{fig:NZ-t-weak}(b) and \ref{fig:NZ-t-strong}(b).
For $\zeta = 0.4$ and $\zeta = 1.0$, the fit exponents $\beta$ are
0.40 and 0.68, respectively, quantifying the larger stretching
for the smaller elongation.
The transient dynamics for intermediate and large $q$ is not
very sensitive to changes of $\zeta$.
Both Figs.~\ref{fig:NZ-t-weak}(b) and \ref{fig:NZ-t-strong}(b)
for $q = 9.8$ demonstrate an initial decay of the correlators 
by about 20\% if the time increases up to $t \approx 0.1$.
This similarity is also reflected by the similarity of the
microscopic time scale $t_{0}$ for the two elongations
listed in Table~\ref{table:1}.
However, for $\epsilon = - 10^{-3}$ or $- 10^{-4}$,
the correlators $\phi_{q,s}^{N}(t)$ require about 100 times
longer times $t$ for $\zeta = 0.4$ than for $\zeta = 1.0$
to decay to zero.
All the enhanced stretching features reflect the fact that
the anomalous exponents $a$ and $b$ are smaller for $\zeta = 0.4$
than for $\zeta = 1.0$ ({\em cf}. Table~\ref{table:1}). 

\begin{figure}
\includegraphics[width=0.8\linewidth]{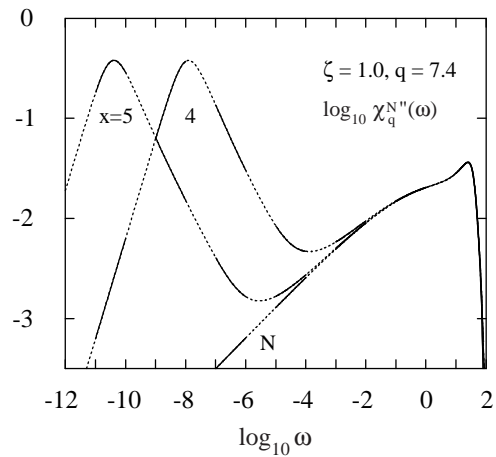}
\caption{
Double logarithmic presentation of susceptibility spectra
$\chi_{q}^{N''} (\omega)$ as function of frequency $\omega$
for the density fluctuations discussed in
Fig.~\protect\ref{fig:NZ-t-strong}(a) (see text).}
\label{fig:susceptibility}
\end{figure}

A comment concerning the accuracy of the numerical solutions
of the equations of motion might be in order.
The primary work consists of calculating $\phi_{q}^{N}(t)$
from Eqs.~(\ref{eq:GLE-N}) and (\ref{eq:MCT-N}) on a 
grid of times.
In the work reported here, 
the initial part of the grid consists of 100 values 
with the equal step size $\delta t = 10^{-5}$.
This interval is then extended by successively
doubling its length and the step size $\delta t$.
By inspection one checks $|\, \phi_{q}^{N}(t) \,| \le 1$
so that the Laplace transform 
$\phi_{q}^{N}(z) = i \int_{0}^{\infty} dt \, \exp(izt) \phi_{q}^{N}(t)$
exists as analytic function for $\mbox{Im } z > 0$.
One checks for $\epsilon < 0$ that $\phi_{q}^{N}(t)$ decreases fast
enough for large $t$ so that
$\phi_{q}^{N}(\omega) = 
\lim_{\eta \to 0} \phi_{q}^{N}(\omega + i \eta) =
\phi_{q}^{N'}(\omega) + i \phi_{q}^{N''}(\omega)$
exists with a smooth reactive part $\phi_{q}^{N'}(\omega)$
and a smooth non-negative spectrum $\phi_{q}^{N''}(\omega)$.
The Fourier integrals are evaluated with a simplified Filon 
procedure~\cite{Tuck67}.
The solid lines in Fig.~\ref{fig:susceptibility} show results 
for the susceptibility spectrum
$\chi_{q}^{N''}(\omega) = \omega \phi_{q}^{N''}(\omega)$
for $\zeta = 1.0$, $q = 7.4$, and
$\epsilon = (\varphi - \varphi_{c}(\zeta))/\varphi_{c}(\zeta) = - 10^{-x}$
with $x = 4,5$, and $\epsilon = 0$.
The correlators are used to evaluate the polynomials
$m_{q}^{N}(t) = \sum_{k,p} V_{q,kp} \phi_{k}^{N}(t) \phi_{p}^{N}(t)$
and from here one gets
$m_{q}^{N}(\omega) = m_{q}^{N'}(\omega) + i m_{q}^{N''}(\omega)$.
Equations~(\ref{eq:GLE-N}) and (\ref{eq:MCT-N}) are solved if
$\phi_{q}^{N}(\omega)$ agrees with 
$\tilde{\phi}_{q}^{N}(\omega) = 
- 1 / \{
\omega - (\Omega_{q}^{N})^{2} / [ \omega + (\Omega_{q}^{N})^{2} m_{q}^{N}(\omega)]
\}.$
Because of causality, it suffices to check
$\tilde{\phi}_{q}^{N''}(\omega) = \phi_{q}^{N''}(\omega)$.
The functions $\omega \tilde{\phi}_{q}^{N''}(\omega)$ are shown as dotted lines
in Fig.~\ref{fig:susceptibility}.
The desired identity is verified by inspection within the
accuracy better than that of the drawing.
Corresponding statements hold for the other functions like
$\phi_{q,s}^{N,Z}(t)$. 
The evaluation of $\phi_{q}^{N}(t)$ on the discrete set of times
from Eqs.~(\ref{eq:GLE-N}) and (\ref{eq:MCT-N})
is done by an extension of the method discussed in 
Ref.~\onlinecite{Goetze96}. 
But it should be noticed that the described proof of the accuracy of the
solution does not require an account of how
the solution $\phi_{q}^{N}(t)$ has been obtained.

\subsection{Dipole correlators}
\label{subsec:3B}

\begin{figure}
\includegraphics[width=0.8\linewidth]{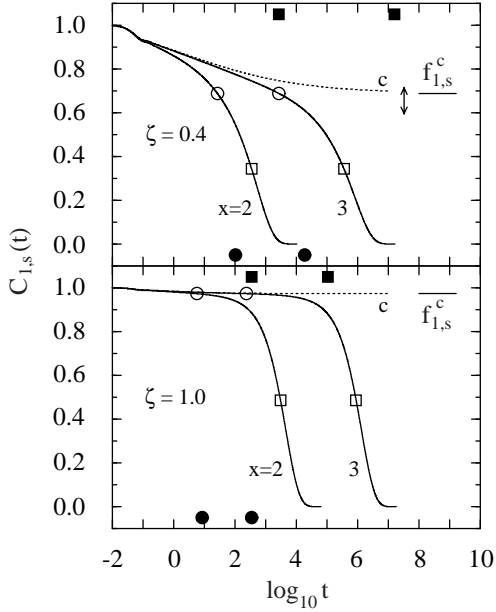}
\caption{
Dipole correlators $C_{1,s}(t)$
for two elongations $\zeta = 0.4$ (upper panel) and $1.0$ (lower panel).
The correlators at the
critical packing fraction $\varphi = \varphi_c$ are shown as
dotted lines marked with $c$.
The plateau $f_{1,s}^{c}$ for each elongation is marked by a
horizontal line.
The distance parameters are chosen as
$\epsilon = (\varphi - \varphi_c) / \varphi_c = - 10^{-x}$ with
$x = 2$ (faster decay) and $x = 3$ (slower decay).
The filled circles and squares mark the
corresponding time scales $t_\sigma$ and $t_\sigma^\prime$,
respectively,
defined in Eqs.~(\protect\ref{eq:first-c}) and
(\protect\ref{eq:t-sigma-prime}).
The open circles and squares on the curves mark the characteristic
time scales $\tau_{\beta}^{1,s}$ and $\tau_{\alpha}^{1,s}$
defined by $C_{1,s}(\tau_{\beta}^{1,s}) = f_{1,s}^{c}$ and
$C_{1,s}(\tau_{\alpha}^{1,s}) = f_{1,s}^{c} / 2$,
respectively.
The vertical line for $\zeta = 0.4$ indicates the decay interval
described by the asymptotic formulas for the $\beta$ process
(see text).}
\label{fig:C1s}
\end{figure}

The dipole correlator of the tagged molecule is defined by
\be
C_{1,s}(t) = \langle {\vec e}_{s}(t) \cdot {\vec e}_{s}(0) \rangle.
\ee
Here, the unit vector ${\vec e}_{s}$ denotes the tagged molecule's
axis. 
$C_{1,s}(t)$ is the reorientational-correlation function 
for angular-momentum index $\ell = 1$.
For a similar reasoning as presented in the paragraph preceding
Eq.~(\ref{eq:phi-N-def}),
it is identical to the coherent reorientational function.
It can be obtained as the zero-wave-vector limit of the charge correlator:
$\phi_{q,s}^{Z}(t) = C_{1,s}(t) + O(q^{2})$~\cite{Chong01}.
$C_{1,s}(t)$ is evaluated most efficiently as follows.
One carries out the $q=0$ limit in Eq.~(\ref{eq:GLE-sNZ}) for $x = Z$
to get the exact Zwanzig-Mori equation
\bea
& &
\partial_{t}^{2} C_{1,s}(t) + 2 v_{R}^{2} \, C_{1,s}(t)
\nonumber \\
& & \quad \,\,
+ \, 
2 v_{R}^{2}
\int_{0}^{t} dt' \, m_{s}^{Z}(t-t') \, \partial_{t'} C_{1,s}(t') = 0,
\label{eq:GLE-C1}
\eea
to be solved with the initial condition
$C_{1,s}(t) = 1 - (v_{R}t)^{2} + O(|t|^{3})$.
The kernel is the $q=0$ limit of the relaxation kernel 
$m_{q,s}^{Z}(t)$ from Eqs.~(\ref{eq:MCT-sNZ}):
\be
m_{s}^{Z}(t) = 
(\zeta^{2} / 72 \pi^{2}) 
\int_{0}^{\infty} dk \,
k^{4} \, \rho S_{k}^{N} \, (c_{k}^{N})^{2} \, w_{k}^{Z} \, 
\phi_{k,s}^{Z}(t) \, \phi_{k}^{N}(t).
\label{eq:MCT-C1}
\ee
The integral is discretized to a sum over $M$ terms as explained 
in connection with Eq.~(\ref{eq:MCT-N-b}). 
Substituting the correlators $\phi_{k,s}^{Z}(t)$ and $\phi_{k}^{N}(t)$,
the kernel $m_{s}^{Z}(t)$ is determined. 
The remaining linear integro-differential equation~(\ref{eq:GLE-C1})
is integrated to yield the desired result for $C_{1,s}(t)$.
Equation~(\ref{eq:MCT-C1}) yields directly the nonergodicity
parameter of the kernel,
$\mu_{s}^{Z} = m_{s}^{Z}(t \to \infty)$,
as integral over the products of $f_{k,s}^{Z} f_{k}^{N}$.
From $\mu_{s}^{Z}$, one derives the probability for the
arrest of the dipole 
$f_{1,s} = C_{1,s}(t \to \infty) = \mu_{s}^{Z} / (1 + \mu_{s}^{Z})$.
This number can also be obtained as
$f_{1,s} = \lim_{q \to 0} f_{q,s}^{Z}$. 
The critical value $f_{1,s}^{c}$ and the corresponding critical
amplitude $h_{1,s} = \lim_{q \to 0} h_{q,s}^{Z}$ were discussed
in Fig.~13 of Ref.~\onlinecite{Chong-MCT-dumbbell-1}.

\begin{figure}
\includegraphics[width=0.8\linewidth]{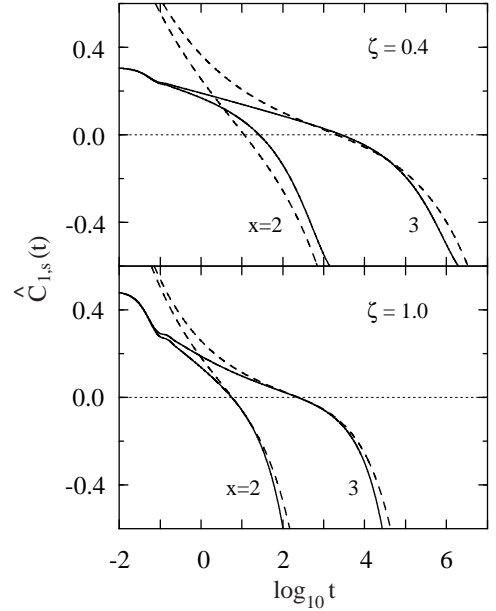}
\caption{
Rescaled dipole correlators
$\hat{C}_{1,s}(t) = [C_{1,s}(t) - f_{1,s}^{c}] / h_{1,s}$
(full lines) for two distance parameters
$\epsilon = (\varphi - \varphi_c) / \varphi_c = - 10^{-x}$ with
$x = 2$ (faster decay) and $x = 3$ (slower decay).
The dashed lines show the $\beta$-correlators
$G(t)$ from Eq.~(\protect\ref{eq:first-b})
for the exponents
$\lambda = 0.885$ and $0.739$ for the elongations
$\zeta = 0.4$ and $1.0$, respectively
({\em cf.} Table~\protect\ref{table:1}).
Here $f_{1,s}^{c} = 0.69$ and 0.97, and
$h_{1,s} = 1.0$ and 0.056
for $\zeta = 0.4$ and 1.0, respectively.}
\label{fig:C1s-beta}
\end{figure}

The dipole correlators $C_{1,s}(t)$ for the elongations
$\zeta = 0.4$ and $1.0$ are 
shown in Fig.~\ref{fig:C1s} for the critical point $\varphi = \varphi_c$ and
for two liquid states.
The time scales $\tau_{\beta}^{1,s}$ and $\tau_{\alpha}^{1,s}$
characterizing the center of $\beta$- and $\alpha$-relaxation processes,
defined by $C_{1,s}(\tau_{\beta}^{1,s}) = f_{1,s}^{c}$ and 
$C_{1,s}(\tau_{\alpha}^{1,s}) = f_{1,s}^{c}/2$, 
are marked by open circles and squares, respectively.
The curves do not clearly exhibit the typical two-step-relaxation 
scenario.
For $\zeta = 1.0$, the critical plateau is so large, 
$f_{1,s}^{c} = 0.97$, that only 3\% of the decay are left for the
transient motion and the critical relaxation.
The results for $\zeta = 0.4$ are influenced by
the nearby type-$A$ transition.
Let us consider the $\beta$-relaxation process for the dipole
dynamics.
The factorization theorem,
Eq.~(\ref{eq:first-a}),
specializes to 
\be
C_{1,s}(t) = f_{1,s}^{c} + h_{1,s} \, G(t).
\label{eq:C1s-first}
\ee
This means that the rescaled correlators
$\hat{C}_{1,s}(t) = [C_{1,s}(t)-f_{1,s}^{c}]/h_{1,s}$ are
given by the $\beta$-correlator $G(t)$.
The latter obeys the scaling law, specified by 
Eq.~(\ref{eq:first-b}) and (\ref{eq:first-c}). 
For fixed rescaled time, $\hat{t} = t/t_{\sigma}$,
the cited formulas deal with the results correctly up to 
order $\sqrt{\sigma}$. 
The leading corrections are of order $|\, \sigma \,|$, and they
explain the range of validity of the leading results~\cite{Franosch97}.
Figure~\ref{fig:C1s-beta} demonstrates the test of the
$\beta$-scaling law. 
On a 10\% accuracy level, the leading-order asymptotic law
accounts for 12\% of the decay of $C_{1,s}(t)$ around the 
plateau for $\zeta = 0.4$, while it describes only 
1.4\% of the decay for $\zeta = 1.0$.
The latter feature is due to the large critical nonergodicity
parameter $f_{1,s}^{c}$, and hence, due to the small critical amplitude $h_{1,s}$, 
for the strong steric hindrance. 
The decay interval described by the leading-order asymptote 
for $\zeta = 0.4$ is indicated in 
Fig.~\ref{fig:C1s} by the vertical line.
For $\epsilon = -0.001$, the corresponding dynamical window
extends from about $t = 1.1 \times 10^{3}$ to about $2.4 \times 10^{4}$, while
it extends from about $t=20$ to $t=70$ for 
$\epsilon = -0.01$.
This discussion requires a reservation. 
The corrections to the scaling results can lead to 
an offset of the plateau~\cite{Franosch97}.
One recognizes that incorporation of such offset would improve the
agreement between numerical solution and its asymptotic description
in the upper panel of Fig.~\ref{fig:C1s-beta}.
The mentioned windows are obtained only after a plateau offset
had been eliminated. 

\begin{figure}
\includegraphics[width=0.8\linewidth]{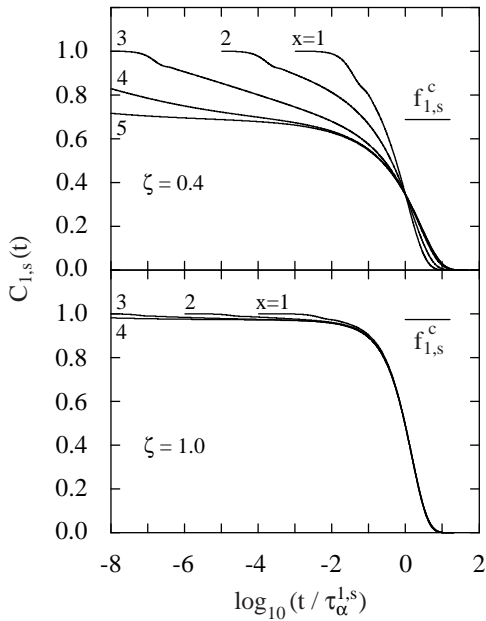}
\caption{
Dipole correlators $C_{1,s}(t)$ for the elongations
$\zeta = 0.4$ (upper panel) and
$1.0$ (lower panel)
for various distance parameters
$\epsilon = (\varphi - \varphi_c) / \varphi_c = - 10^{-x}$
with $x$ specified in the figure,
presented as function of $\log_{10}(t/\tau_{\alpha}^{1,s})$.
The $\alpha$-relaxation-time scale $\tau_{\alpha}^{1,s}$ is
defined by $C_{1,s}(\tau_{\alpha}^{1}) = f_{1,s}^{c}/2$.
The horizontal lines indicate the plateaus $f_{1,s}^{c}$.}
\label{fig:C1s-alpha}
\end{figure}

The $\alpha$-relaxation scaling law for the dipole correlator
reads according to Eq.~(\ref{eq:phi-Y-second}):
\be
C_{1,s}(t) = \tilde{C}_{1,s}(\tilde{t}), \quad
\tilde{t} = t/t_{\sigma}^{\prime}, \quad
|\, \sigma \,| \ll 1, \quad
t_{\sigma} \ll t.
\ee
The $\epsilon$-independent master function $\tilde{C}_{1,s}$
is determined by the $q=0$ limit of the mode-coupling functional
${\cal F}_{q,s}^{Z}$ given in Eq.~(\ref{eq:MCT-sNZ-b})
at the critical point. 
The $\alpha$-time scale $\tau_{\alpha}^{1,s}$ 
can be defined by $C_{1,s}(\tau_{\alpha}^{1,s}) = f_{1,s}^{c}/2$,
and shall be written as 
$\tau_{\alpha}^{1,s} = \tilde{t}_{1,s} \, t_{\sigma}^{\prime}$.
Here, the $\alpha$-scale factor $\tilde{t}_{1,s}$ is defined 
in terms of the master function as
$\tilde{C}_{1,s}(\tilde{t}_{1,s}) = f_{1,s}^{c}/2$.
The scaling law implies that a representation of $C_{1,s}(t)$
as a function of the rescaled time $t/\tau_{\alpha}^{1,s}$
should superimpose correlators for different distance parameters
$\epsilon$ on the common curve
$\tilde{C}_{1,s}(\tilde{t}/\tilde{t}_{1,s})$.
Asymptotic validity means that the $\log(t/\tau_{\alpha}^{1,s})$
interval, where the scaling law is obeyed, expands to an arbitrary size
for $\epsilon \to 0$.

The lower panel of Fig.~\ref{fig:C1s-alpha} demonstrates that 
the described scenario
for the evolution of the $\alpha$ process is valid for
the elongation $\zeta = 1.0$.
On the other hand, 
the upper panel shows that 
the dipole correlators for $\zeta = 0.4$
exhibit the superposition principle
only for $|\, \epsilon \,| \le 10^{-4}$.
In particular, the plateau region emerges only for these extremely 
small values of the distance 
parameter $|\, \epsilon \,|$. 
As discussed in Ref.~\onlinecite{Franosch97}, this is because of the
correction to the leading-order asymptotic law,
which reads for not too large values of rescaled time $\tilde{t}$:
\be
C_{1,s}(t) = \tilde{C}_{1,s}(\tilde{t}) + 
|\sigma| B_{1} h_{1,s}  / \tilde{t}^{b},
\label{eq:C1s-alpha-correction}
\ee
with $B_{1} = \frac{1}{2} / [\Gamma(1-b) \Gamma(1+b) - \lambda]$.
Therefore, the correction is larger, 
the larger the product $B_{1} h_{1,s}$ is. 
As discussed in connection with Fig.~13 of 
Ref.~\onlinecite{Chong-MCT-dumbbell-1},
the size of the critical amplitude $h_{1,s}$ becomes very large
near the critical elongation $\zeta_{c}$.
In addition, $B_{1}$ increases with $\lambda$.
One gets $B_{1} = 1.56$ and 0.45 for $\zeta = 0.4$ and 1.0,
respectively. 
Thus, the anomaly shown in the upper panel of 
Fig.~\ref{fig:C1s-alpha} is due to the large correction to the
leading-order asymptotic law, caused by the precursor
effect of the nearby type-$A$ transition and by the large value 
for $\lambda$.
The descriptions of correlators by the first and second 
scaling laws overlap for $t \approx t_{\sigma}$. 
Therefore, the anomaly for $\zeta = 0.4$ exhibited in 
Fig.~\ref{fig:C1s-alpha} can also be explained as 
the large percentage of the decay of $C_{1,s}(t)$
described by the $\beta$-scaling law 
({\em cf.} Fig.~\ref{fig:C1s}).

\subsection{Mean-squared displacements}
\label{subsec:3C}

There are two mean-squared displacements (MSD)
to be considered for the symmetric dumbbell.
One refers to the position of the constituent atom of the
tagged molecule ${\vec r}_{s}^{\, A}$, and the other to the
molecule's center 
${\vec r}_{s}^{\, C} = ({\vec r}_{s}^{\, A} + {\vec r}_{s}^{\, B}) / 2$:
\be
\Delta_{A,C}(t) = 
\langle [ {\vec r}_{s}^{\, A,C}(t) - {\vec r}_{s}^{\, A,C}(0) ]^{2} \rangle \, / \, 6.
\ee
Here, a factor 6 is introduced in the definition for later convenience.
Since there is the relation~\cite{Chong01}
\be
\Delta_{A}(t) = \Delta_{C}(t) + (\zeta^{2}/12) \, [1 - C_{1,s}(t)],
\label{eq:Delta-A-decom}
\ee
it is sufficient to calculate $\Delta_{C}(t)$.
This function is determined by the small-$q$ limit of the
density correlator: 
$\phi_{q,s}^{N}(t) = 1 - q^{2} \Delta_{C}(t) + O(q^{4})$~\cite{Chong01}.
The small-$q$ expansion of Eq.~(\ref{eq:GLE-sNZ}) for $x = N$ leads to
the exact equation of motion
\begin{equation}
\partial_{t}^{2} \Delta_{C}(t) - v_{T}^{2} + v_{T}^{2}
\int_{0}^{t} dt' \, m_{s}^{N}(t-t') \, 
\partial_{t'} \Delta_{C}(t') = 0,
\label{eq:GLE-MSD}
\end{equation}
to be solved with the initial behavior 
$\Delta_{C}(t) = (1/2) (v_{T} t)^2 + O (|t|^3)$. 
The kernel is obtained as the $q=0$ limit of 
Eq.~(\ref{eq:MCT-sNZ-a}) for $x = N$:
\be
m_{s}^{N}(t) = 
(1 / 6 \pi^{2}) 
\int_{0}^{\infty} dk \,
k^{4} \, \rho S_{k}^{N} \, (c_{k}^{N})^{2} \, w_{k}^{N} \, 
\phi_{k,s}^{N}(t) \, \phi_{k}^{N}(t).
\label{eq:MCT-MSD}
\ee
After the discretization of the integral as explained above
and substituting the results for the two density correlators,
the kernel $m_{s}^{N}(t)$ is determined.
Then, the linear equation~(\ref{eq:GLE-MSD}) is integrated to
yield the MSD of the molecule's center. 

The long-time behavior of the MSD 
depends sensitively on control parameters 
near the transition singularity, and therefore it is well suited
to study the glass-transition precursors.
In liquid states, one obtains
from Eq.~(\ref{eq:GLE-MSD}) the long-time asymptote: 
$\lim_{t \to \infty} \Delta_{C} (t) / t = D$.
Here $D$ is the diffusion constant, and it is expressed
as the inverse of the zero-frequency spectrum of the relaxation
kernel: 
$D = 1 / \int_{0}^{\infty} dt \, m_{s}^{N}(t)$.
In glass states, the tagged molecule's total density fluctuations 
arrest for long times:
$\phi_{q,s}^{N} (t \to \infty) = f_{q,s}^{N} > 0$. 
The Lamb-M\"ossbauer factor $f_{q,s}^{N}$
approaches unity for $q$ tending to zero, and
a localization length, $r_{C}$, of the center characterizes the
width of the $f_{q,s}^{N}$--versus--$q$ curve: 
$f_{q,s}^{N} = 1 - (q r_{C})^{2} + O(q^4)$. 
One gets $\lim_{t \to \infty} \Delta_{C} (t) = r_{C}^{2}$. 
Using Eq.~(\ref{eq:GLE-MSD}), 
one can express $r_{C}^{2}$ as inverse of the long-time limit of
the relaxation kernel from Eq.~(\ref{eq:MCT-MSD}): 
$r_{C}^{2} = 1 / m_{s}^{N}(t \to \infty)$. 

\begin{figure}
\includegraphics[width=0.8\linewidth]{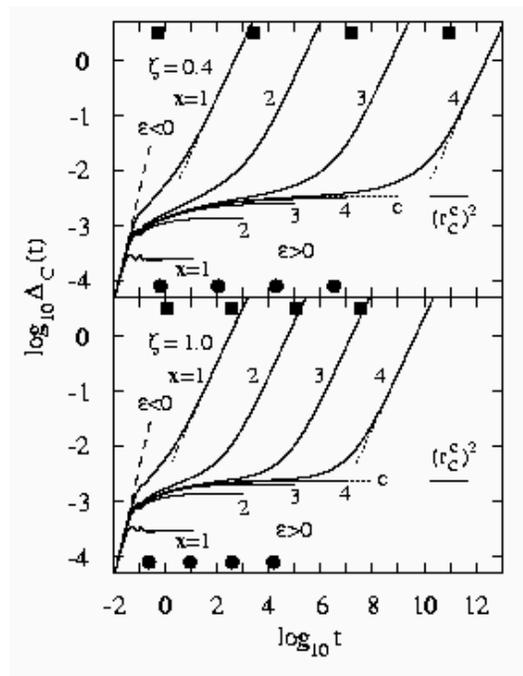}
\caption{Double logarithmic presentation of
the mean-squared displacement for the center of mass
$\Delta_{C}(t)$ for the elongations $\zeta = 0.4$ (upper panel)
and $1.0$ (lower panel).
The dotted lines with label $c$
refer to the critical packing fraction $\varphi_c$, and the solid lines
to $\epsilon = (\varphi - \varphi_c) / \varphi_c = \pm 10^{-x}$
for $x = 1$, 2, 3, 4.
The straight dashed line with slope 2 in each panel exhibits the
ballistic asymptote $(v_{T} t)^2 /2$.
The straight dotted lines with
slope 1 denote the long-time asymptotes $Dt$ of the two
liquid curves for $x = 1$ and $4$.
The horizontal lines
mark the square of the critical localization lengths
$(r_{C}^{c})^{2}$.
The filled circles and squares mark the characteristic
times $t_{\sigma}$ and $t_{\sigma}^{\prime}$ , respectively,
defined in
Eqs.~(\protect\ref{eq:first-c}) and (\protect\ref{eq:t-sigma-prime})
for $x = 1,2,3,4$.}
\label{fig:MSD-COM}
\end{figure}

The ideal liquid-glass transition implies a transition from a
regime with molecule's diffusion for $\varphi < \varphi_c$ to one
with molecule's localization for $\varphi \geq \varphi_c$. 
The former is characterized by $D > 0$ and $1 / r_{C} = 0$, 
and the latter by $D = 0$ and $1 / r_{C} > 0$. 
The subtleties of the
glass-transition dynamics occur outside
the transient regime. 
Figure~\ref{fig:MSD-COM}
exhibits the MSD of the molecule's center for various
packing fractions near the transition. 
For very short times, say $t \leq t_0$, interaction effects are unimportant
and $\lim_{t \to 0} \Delta_{C} (t) /t^2 = v_{T}^2 / 2$
reflects ballistic motion. 
For times larger than $t_0$, the
cage effect leads to a suppression of $\Delta_{C} (t)$ below the
short-time asymptote. 
For long times in liquid states, the MSD approaches the diffusion
asymptote, 
$\lim_{t \to \infty} \Delta_{C} (t) / t = D$, as shown by the dotted
straight lines drawn 
for the curves with labels $x = 1$ and $x = 4$. 
Upon increasing $\varphi$ towards $\varphi_c$, the
diffusivity decreases towards zero. 

\begin{figure}
\includegraphics[width=0.7\linewidth]{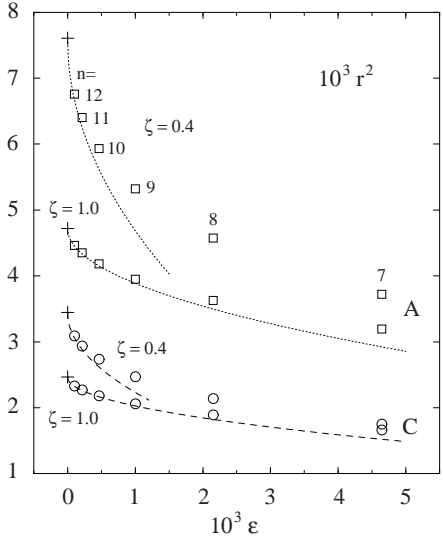}
\caption{Square of the localization lengths for
the molecule's center $r_{C}^{2}$ (circles) and
for the constituent atom $r_{A}^{2}$ (squares) for
$\epsilon = (\varphi - \varphi_{c})/\varphi_{c} = 10^{-n/3}$,
$n = 7,8,\cdots,12$.
The crosses mark the critical values $(r_{C}^{c})^{2}$ and
$(r_{A}^{c})^{2}$, respectively.
The dashed and dotted lines are the leading order asymptotic laws
$r_{X}^{2} = (r_{X}^{c})^{2} - h_{X} \sqrt{\sigma/(1-\lambda)}$
for $X=C$ and $A$, respectively.}
\label{fig:rC-rA-vs-epsilon}
\end{figure}

The curves with $x = 1$ for $\epsilon > 0$ in 
Fig.~\ref{fig:MSD-COM}
deal with the glass state 
$\varphi = 1.1 \varphi_c$.
For this density, there is no obvious glassy dynamics.
Rather, $\Delta_{C} (t)$ has approached its long-time limit $r_{C}^2$
after the oscillations have disappeared for $t \approx 1$.
Decreasing $\varphi$ towards $\varphi_c$, the softening of the
glass manifests itself by an increase of the localization length $r_{C}$. 
At the transition point $\varphi = \varphi_c$, the critical value
$r_{C}^{c} = 0.0587$ ($0.0497$)
for $\zeta = 0.4$ ($1.0$) is reached. 
Substituting Eq.~(\ref{eq:fq}) for $f_{q}^{N}$ and 
the analogous formula for $f_{q,s}^{N}$
to evaluate $r_{C}^{2}$ from Eq.~(\ref{eq:MCT-MSD}),
it follows that the glass
instability at $\varphi_c$ causes a $\sqrt{\sigma}$--anomaly
for the localization length,
\begin{equation}
r_{C}^{2} = (r_{C}^{c})^{2} - h_{C} \, 
\sqrt{\sigma / (1 - \lambda)} + O(\sigma),
\label{eq:rC-2}
\end{equation}
where 
$h_{C} = 9.66 \times 10^{-3}$ ($5.14 \times 10^{-3}$)
for the elongation $\zeta = 0.4$ ($1.0$). 
A corresponding formula holds for the square of the localization
length $r_{A}$ for the constituent atom
with $(r_{C}^{c})^{2}$ replaced by
$(r_{A}^{c})^{2} = (r_{C}^{c})^{2} + (\zeta^{2}/12) (1 - f_{1,s}^{c})$
and $h_{C}$ replaced by 
$h_{A} = h_{C} + (\zeta^{2}/12) h_{1,s}$
according to Eq.~(\ref{eq:Delta-A-decom}).
In Fig.~\ref{fig:rC-rA-vs-epsilon}, 
the leading order results for $r_{C}^{2}$ and $r_{A}^{2}$
are shown as dashed and dotted lines, respectively.
For $\zeta = 1.0$, the data are described by the
square-root law for $\epsilon \le 2 \times 10^{-3}$.
Similar small intervals for the validity of the leading order
descriptions have been found for the Debye-Waller factor of the
HSS for small wave vector $q$~\cite{Franosch97}.
For $\zeta = 0.4$, the range of validity of the universal formula
is reduced to the even smaller intervals
$\epsilon < 0.5 \times 10^{-3}$.

The glass curves for $\epsilon = 0.01$, shown in 
Fig.~\ref{fig:MSD-COM} with label
$x = 2 $, exhibit a decay between the end of the transient
oscillations and the arrest at $r_{C}^2$ which is stretched over a
time interval of about two orders of magnitude.
A similar two-decade interval is needed for the liquid curves 
($\epsilon < 0$) with label 
$x = 2$ to reach the critical value $(r_{C}^{c})^{2}$.
After crossing $(r_{C}^{c})^{2}$, two further decades of an upward bent
$\log_{10} \Delta_{C} (t)$--versus--$\log_{10} t$ variation is
exhibited before the final diffusion asymptote is reached. 
The stretching is more enhanced as the system is driven
towards the transition point, $| \, \epsilon \, | \to 0$.
The indicated slow and stretched time variation outside
the transient regime is the glassy
dynamics exhibited by the MSD.
Again, the stretching is more pronounced for $\zeta = 0.4$
than for $\zeta = 1.0$. 

The first scaling law deals with the dynamics where
$\Delta_{C}(t) - (r_{C}^{c})^{2}$ is small.
One gets in analogy to Eqs.~(\ref{eq:first-a}) and (\ref{eq:first-b}):
\be
\Delta_{C}(t) = (r_{C}^{c})^{2} - h_{C} \, \sqrt{|\, \sigma \,|} \,
g_{\pm}(t/t_{\sigma}), \quad \sigma \, \gtrless\,  0,
\label{eq:MSD-first-a}
\ee
for $|\, \sigma \,| \, \ll 1$ and $t \gg t_{0}$. 
For glass states with $\sigma > 0$, 
this formula describes the approach
towards the arrest at $r_{C}^{2} = \Delta_C (t \to \infty)$.
For liquid states with $\sigma < 0$,
it describes how the 
$\Delta_{C}(t)$--versus--$t$ curve crosses and leaves
the plateau $(r_{C}^{c})^{2}$.
In particular, one gets von Schweidler's law for large
$t/t_{\sigma}$ in analogy to Eq.~(\ref{eq:phi-Y-von}):
\begin{equation}
\Delta_{C}(t) = (r_{C}^{c})^{2} + h_{C} \, (t/t_{\sigma}^{\prime})^{b},
\quad \sigma \to -0, \quad t_{\sigma} \ll t \ll t_{\sigma}^{\prime}. 
\label{eq:MSD-first-b}
\end{equation}
The increase of $\Delta_{C}(t)$ above the plateau towards
the diffusion asymptote is the $\alpha$ process of the MSD. 
In analogy to Eq.~(\ref{eq:phi-Y-second}), there holds
the superposition principle
\begin{equation}
\Delta_{C}(t) = \tilde{\Delta}_{C}(\tilde{t}), \quad
\tilde{t} = t/t_{\sigma}^{\prime}, \quad 
|\, \sigma \,| \, \ll 1, \quad t_{\sigma} \ll t.
\label{eq:MSD-second}
\end{equation}
In Ref.~\onlinecite{Chong01b}, it has been discussed in detail
how these leading-order asymptotic results 
can account quantitatively for the glassy dynamics of the MSD,
albeit for the tagged spherical particle and the dumbbell molecule
immersed in the hard-sphere system.
Instead of repeating such an extensive analysis, 
we shall only show here some representative examples. 

\begin{figure}
\includegraphics[width=0.8\linewidth]{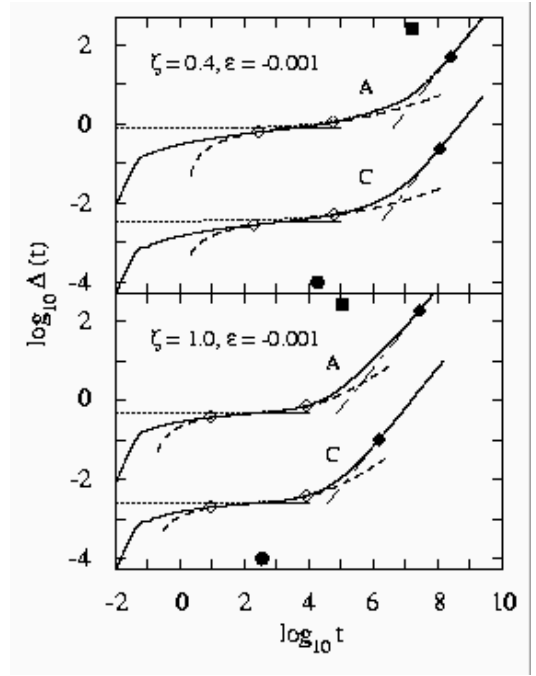}
\caption{
Double logarithmic presentation of the mean-squared displacement
for the constituent atom $\Delta_{A}(t)$ and
for the center $\Delta_{C}(t)$
for the elongations $\zeta = 0.4$ (upper panel)
and $\zeta = 1.0$ (lower panel).
The curve $\Delta_{A}(t)$ is shifted upwards by two
decades to avoid overcrowding.
The distance parameter is $\epsilon = - 10^{-3}$,
and the corresponding times $t_\sigma$ and
$t_\sigma^\prime$ are marked by filled
circles and squares, respectively.
The dashed lines are the first-scaling-law results,
Eq.~(\protect\ref{eq:MSD-first-a}).
The open diamonds mark the points where the dashed lines differ
from the solid ones by 5\%.
The straight dashed-dotted lines
exhibit the diffusion asymptotes $Dt$,
and the filled diamonds mark the position where
these differ from the solid lines by 5\%.
The dotted lines exhibit the
second-scaling-law results, Eq.~(\ref{eq:MSD-second}).
The horizontal lines mark the square of the localization lengths
$(r_{A}^{c})^{2}$ and $(r_{C}^{c})^{2}$.}
\label{fig:MSD-scaling}
\end{figure}

The upper and lower panels of Fig.~\ref{fig:MSD-scaling}, 
respectively for the elongations $\zeta = 0.4$ and $1.0$, 
exhibit such tests of the scaling-law descriptions 
for a liquid state
which is sufficiently close to the transition point,
$\epsilon = -10^{-3}$.
Also, the MSD of the constituent atom
$\Delta_{A}(t)$ is considered in these figures.
Because of Eq.~(\ref{eq:Delta-A-decom}), 
the asymptotic laws (\ref{eq:MSD-first-a}) and (\ref{eq:MSD-first-b})
for $\Delta_{A}(t)$ 
hold with $(r_{C}^{c})^{2}$ replaced by
$(r_{A}^{c})^{2} = (r_{C}^{c})^{2} + (\zeta^{2}/12) (1 - f_{1,s}^{c})$
and $h_{C}$ replaced by 
$h_{A} = h_{C} + (\zeta^{2}/12) h_{1,s}$.
The range of validity of the first-scaling-law description is
indicated by the open diamonds. 
One finds that such a range is nearly the same for both 
$\Delta_{C}(t)$ and $\Delta_{A}(t)$,
and this holds
irrespective of the molecule's elongation $\zeta$. 
The beginning of the $\alpha$-process of the MSD, i.e., 
its initial increase above the plateau, is described by
von Schweidler's law, Eq.~(\ref{eq:MSD-first-b}).
It is exhibited for $t \gg t_{\sigma}$ by the dashed lines.
The $\alpha$-process terminates in the diffusion law for long times,
exhibited by the straight dashed-dotted lines.
The beginning of the diffusion law is indicated by the filled
diamonds for each MSD curve. 
The $\alpha$-process follows well the second scaling law,
Eq.~(\ref{eq:MSD-second}), which is 
presented by the dotted lines.
The descriptions by the two scaling laws overlap for
$t \approx t_{\sigma}$. 
Together, the two scaling laws provide a description of the glassy
dynamics of the MSD for $t \gtrsim 100$ and $t \gtrsim 10$ for
$\zeta = 0.4$ and $\zeta = 1.0$, respectively.
Notice that there is a large interval of times outside the
transient regime, say $\log_{10} t \gtrsim -0.5$, where the
structural relaxation is not described by the first scaling law.
This is a peculiarity of the MSD~\cite{Chong01b} which is not
found for the other functions discussed here.

The crossover window from the end of the von-Schweidler-law 
regime (indicated by the right open diamonds) to the beginning of the
diffusion process (indicated by the filled diamonds) for the MSD of
the center, $\Delta_{C}(t)$, is about 
3 (2) decades wide for $\zeta = 0.4$ ($1.0$).
On the other hand, 
the corresponding window for the MSD of the constituent atom,
$\Delta_{A}(t)$, 
is larger than that of $\Delta_{C}(t)$ by a factor of
about 7 (13) for $\zeta = 0.4$ ($1.0$).
As discussed in Ref.~\onlinecite{Chong01b},
this enlarged crossover window for $\Delta_{A}(t)$ is caused by
the rotation-translation coupling, i.e.,
by the second term on the right-hand side of 
Eq.~(\ref{eq:Delta-A-decom}).
This effect is smaller for $\zeta = 0.4$ because the relaxation
of the second term, determined by that of $C_{1,s}(t)$,
is considerably enhanced due to the nearby type-$A$
transition as discussed in connection with 
Fig.~\ref{fig:C1s}. 

\begin{figure}
\includegraphics[width=0.7\linewidth]{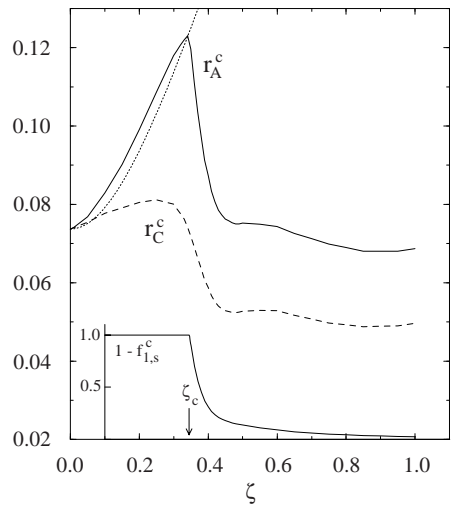}
\caption{
Critical localization lengths for
the constituent atom $r_{A}^{c}$ (solid line) and
the center $r_{C}^{c}$ (dashed line)
along the liquid-glass-transition line
parameterized by the elongation $\zeta$.
The dotted line represents the function
$\sqrt{(r_{C}^{c})^{2} + \zeta^2 / 12}$
discussed in the text.
The inset exhibits $1 - f_{1,s}^{c}$ as function of $\zeta$, where
$f_{1,s}^{c} = C_{1,s}(t \to \infty)$ is the long-time limit of
the dipole correlator $C_{1,s}(t)$.
The arrow indicates the critical elongation $\zeta_{c} = 0.345$.}
\label{fig:rA-rC-vs-zeta}
\end{figure}

Figure~\ref{fig:rA-rC-vs-zeta} exhibits the critical localization lengths
of the molecule's center $r_{C}^{c}$ and of the constituent atom $r_{A}^{c}$
as function of the elongation $\zeta$.
There are three distinct regions: 
one for small elongations, say $\zeta \le 0.3$, 
another for large elongations, say $\zeta \ge 0.4$, and the
crossover region between them. 
For $\zeta \le \zeta_{c}$, 
the localization length of the center $r_{C}^{c}$
is nearly constant while that of the constituent $r_{A}^{c}$
increases rapidly with $\zeta$.
Both $r_{C}^{c}$ and $r_{A}^{c}$ decrease rapidly within the
crossover region, and they decrease only slightly as function of
the elongation for $\zeta \ge 0.4$. 
The rapid increase of $r_{A}^{c}$ for $\zeta \le \zeta_{c}$ 
can be understood as follows.
Since $C_{1,s}(t \to \infty) = 0$ for $\zeta \leq \zeta_c$, one gets
from the long-time limit of Eq.~(\ref{eq:Delta-A-decom}):
$r_A^{c} = \sqrt{(r_{C}^{c})^{2} + \zeta^2 / 12}$.
The dotted line in Fig.~\ref{fig:rA-rC-vs-zeta} 
shows this formula with $r_{C}^{c}$ fixed to the value 
for $\zeta = 0.0$.
It explains the increase of $r_{A}^{c}$
for $\zeta$ increasing up to $\zeta_c$.
Thus, $r_{A}^{c}$ increases because the reorientational 
dynamics of the molecule's axis
is not arrested for $\zeta \le \zeta_{c}$. 
In this sense, the localization of molecules with small elongations
is primarily caused by that of the molecule's center.
This is consistent with the result 
that the glass transition for systems with small elongations 
is mainly driven by the arrest of the 
center-of-mass density fluctuations~\cite{Chong-MCT-dumbbell-1}.
The critical localization length $r_{C}^{c}$ provides 
the upper limit for $r_{C}$ characterizing the size of the arrested
structure, and its value $0.07 \sim 0.08$ for 
$\zeta \le \zeta_{c}$ is consistent with 
Lindemann's melting criterion. 
The critical localization length of the constituent atom $r_{A}^{c}$
for $\zeta \ge 0.4$ has similar values. 
This implies that the localization of molecules with large elongations
is caused by that of the molecule's constituents rather than by
that of the molecule's center:
the localization of the center is subordinate to that of the constituents. 
As explained in Ref.~\onlinecite{Chong-MCT-dumbbell-1}, 
angular correlations become more relevant for the glass transition of systems
with large elongations. 
Such angular correlations result in more efficient localization of the
constituent atoms, 
and this is the reason for the relevance of $r_{A}^{c}$
for the localization of molecules with large elongations.
The center localization is reduced by about $1/\sqrt{2}$,
as expected for independent motion of the two constituents.

\subsection{The {\protect\boldmath $\alpha$}-relaxation scales}
\label{subsec:3D}

\begin{figure}
\includegraphics[width=0.8\linewidth]{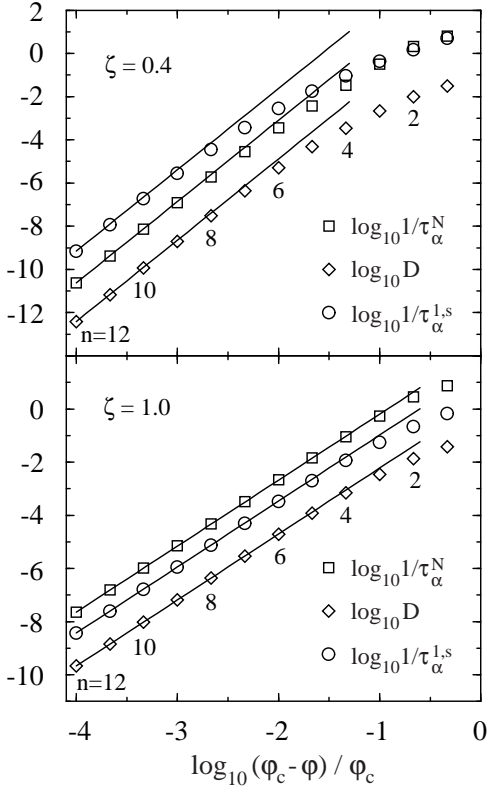}
\caption{
Double logarithmic presentation of the $\alpha$-relaxation rates
$1/\tau_{\alpha}^{N}$ (squares) 
for the coherent total density correlator 
$\phi_{q}^{N}(t)$ at the structure-factor-peak position,
$1/\tau_{\alpha}^{1,s}$ (circles) for the dipole correlator
$C_{1,s}(t)$,
and the diffusion coefficients 
$D$ (diamonds) 
for the elongations $\zeta = 0.4$ (upper panel) and $1.0$ (lower panel)
as function of the reduced packing fraction 
$(\varphi_{c} - \varphi)/\varphi_{c} = 10^{-n/3}$.
The $\alpha$-scaling times $\tau_{\alpha}^{N}$ and 
$\tau_{\alpha}^{1,s}$ 
are defined as in Eq.~(\protect\ref{eq:tau-alpha}).
The solid lines are the power-law asymptotes
$\Gamma_{A} |\,  \epsilon \, |^{\gamma}$ 
(see text) 
with $\gamma = 3.77$ and $2.48$ for $\zeta = 0.4$ and $1.0$,
respectively.
The prefactors $\Gamma_{A}$ have been chosen so that the
solid lines go through the data points for $n = 12$.}
\label{fig:scaling-time}
\end{figure}

The superposition principles for various correlators imply coupling of
the $\alpha$-relaxation time scales or relaxation rates
as described in Sec.~\ref{subsec:2B}. 
This scale coupling or $\alpha$-scale universality is demonstrated in 
Fig.~\ref{fig:scaling-time} for the rate
$1/\tau_{\alpha}^{N}$ of the coherent total density correlator 
$\phi_{q}^{N}(t)$
for the wave number $q \approx 7$,
the rate $1/\tau_{\alpha}^{1,s}$ 
of the dipole correlator $C_{1,s}(t)$,
and the diffusion constant $D$. 
Here, the $\alpha$-relaxation times for $\phi_{q}^{N}(t)$ and
$C_{1,s}(t)$ are determined by the convention given in 
Eq.~(\ref{eq:tau-alpha}).

\begin{figure}
\includegraphics[width=0.7\linewidth]{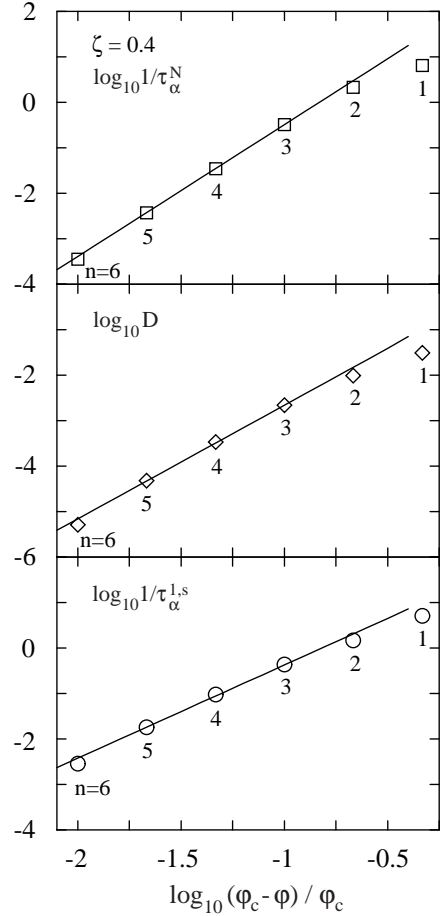}
\caption{
Results as in Fig.~\ref{fig:scaling-time},
but only for a restricted region of the reduced packing fraction,
$n \le 6$, for the elongation $\zeta = 0.4$.
The solid lines here are the power-law fits
$\Gamma_{A}^{\rm eff} | \, \epsilon \, |^{\gamma^{\rm eff}}$ with the effective
power-law exponents
$\gamma^{\rm eff} = 2.90$, $2.50$, and $2.05$ for
$1/\tau_{\alpha}^{N}$, $D$, and $1/\tau_{\alpha}^{1,s}$,
respectively.
The prefactors $\Gamma_{A}^{\rm eff}$ have been chosen so that the
solid lines go through the data points for $n = 5$.}
\label{fig:scaling-time-effective}
\end{figure}

Let us first consider the results for $\zeta = 1.0$.
Although the asymptotic behavior is the same for all the quantities,
$1/\tau_{\alpha}^{1,s}$ and $D$ start to deviate visibly from their asymptotic
results for $n=4$, while $1/\tau_{\alpha}^{N}$ starts to deviate only for
$n=2$.
The mentioned $\alpha$-scale universality holds 
in the leading asymptotic limit for $\sigma \to -0$, and the 
corrections to the asymptotic predictions are different for different 
quantities. 
Thus, the found feature in the results for $\zeta = 1.0$ 
underlines the nonuniversality of the deviations.
Let us add that the range of validity of the asymptotic-law description
is quite similar to that of the hard-sphere 
system~\cite{Franosch97,Fuchs98}.
The results for $\zeta = 1.0$ follow the pattern 
that has been analyzed theoretically so far. 
The results for $\zeta = 0.4$ exhibit much more pronounced deviations 
from the asymptotic-law predictions:
$1/\tau_{\alpha}^{N}$ and $D$ start to deviate visibly for $n=7$, and 
the deviation of $1/\tau_{\alpha}^{1,s}$ starts even for $n=9$.
These deviations are due to the corrections to the $\alpha$-scaling law.
Their magnitudes are proportional to the product of the
critical amplitudes for each quantity
and the coefficient $B_{1}$ as noted in 
Eq.~(\ref{eq:C1s-alpha-correction}). 
The critical amplitude $h_{q}^{N}$ for $\zeta = 0.4$ at the 
structure-factor-peak position is slightly 
larger than the corresponding result
for $\zeta = 1.0$
({\em cf}. Fig.~9 of Ref.~\onlinecite{Chong-MCT-dumbbell-1}).
In addition, $B_{1}$ is 3.5 times larger for $\zeta = 0.4$
than for $1.0$. 
This is the reason for the larger deviation of $1/\tau_{\alpha}^{N}$ found for
$\zeta = 0.4$ than the one for $\zeta = 1.0$, 
and similarly for $D$. 
The even more pronounced deviation for $1/\tau_{\alpha}^{1,s}$
is due to the large critical amplitude $h_{1,s}$ caused by the
precursor effect of the nearby type-$A$ transition, as discussed
in connection with Fig.~13 of Ref.~\onlinecite{Chong-MCT-dumbbell-1}.
Another possible source for the large deviations for $\zeta = 0.4$
might be due to higher-order glass transition singularities, 
which can lead to the violation of the second scaling law,
and thus the $\alpha$-scale universality.
The signature of such singularities is the approach of $\lambda$
towards unity, which implies a divergency of $B_{1}$. 

A remark shall be added concerning the determination of 
the exponent $\gamma$ entering the power-law behavior for the 
$\alpha$-relaxation time scale or relaxation rate 
as specified by Eq.~(\ref{eq:scale-coupling}).  
This result is based on the validity of the 
scaling law. 
Therefore, one cannot appeal to MCT if one fits power laws
for $\alpha$-relaxation rates for cases where the scaling law is violated
so strongly as shown in the upper panel of Fig.~\ref{fig:C1s-alpha}
for $x \le 2$, i.e. for $n \le 6$. 
Figure~\ref{fig:scaling-time-effective} demonstrates that,
for $\zeta = 0.4$, the $\alpha$-relaxation rates
for $| \, \epsilon \, | \ge 10^{-2}$
can be fitted well by power laws
$1/\tau \propto |\, \epsilon \,|^{\gamma^{\rm eff}}$  
for an about 1.3-decade variation of the distance parameter 
$| \, \epsilon \, |$.
The used effective exponent $\gamma^{\rm eff} = 2.90$
(2.50; 2.05)
describes the variation of 
$1/\tau_{\alpha}^{N}$ ($D$; $1/\tau_{\alpha}^{1,s}$) 
over 3.5 (3.3; 2.5) orders of magnitude.
The variable-dependent effective exponent $\gamma^{\rm eff}$
again underlines the nonuniversality of the deviations from the
scaling law. 
The quantities $\gamma^{\rm eff}$ are mere fit parameters
without well-defined meaning for the discussion of our theory.
They depend on the interval of $|\, \epsilon \,|$ chosen for
the fit.

\begin{figure}
\includegraphics[width=0.8\linewidth]{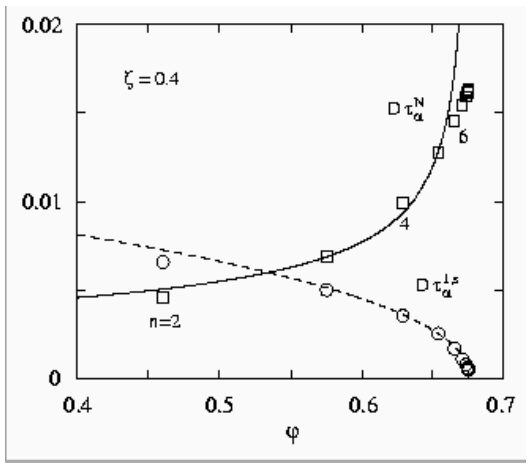}
\caption{
Products $D \tau_{\alpha}^{N}$ (squares) and $D \tau_{\alpha}^{1,s}$
(circles) for the $\alpha$-relaxation scales of the $\zeta = 0.4$ liquid
shown in
Figs.~\protect\ref{fig:scaling-time} and
\protect\ref{fig:scaling-time-effective}.
The full (dashed) line is a power-law fit
$\Gamma^{\prime} |\, \epsilon \,|^{\gamma^{\prime}}$ with
$\Gamma^{\prime}$ chosen so that the line goes through
the data point for $n = 5$ and $\gamma^{\prime} = -0.40$ ($0.45$).}
\label{fig:D-tau}
\end{figure}

Another manner for testing the coupling between
the scales of two variables, say $Y$ and $Y'$, is provided
by a plot of the ratio $\tau_{\alpha}^{Y} / \tau_{\alpha}^{Y'}$
versus the control parameter.
Such a plot does not require the knowledge of $\varphi_{c}$,
nor is it biased by the choice of some fit interval. 
This is demonstrated in Fig.~\ref{fig:D-tau} for the ratios
formed with $\tau_{\alpha}^{N}$ or $\tau_{\alpha}^{1,s}$ and the
scale for the diffusivity $\tau_{\alpha}^{D} \propto 1/D$.
Instead of weakly $\varphi$-dependent ratios expected for the 
asymptotic law, the figure shows variations by more than
a factor 3.
The shown plot suggests power-law fits
$\tau_{\alpha}^{Y} / \tau_{\alpha}^{Y'} \propto 
(\varphi_{c} - \varphi)^{\gamma^{\prime}}$
as shown by the full line with $\gamma^{\prime} = -0.40$ and the
dashed one with $\gamma^{\prime} = 0.45$.
The fit exponents $\gamma^{\prime}$ are the differences of the
effective exponents $\gamma^{\rm eff}$ referring to 
$\tau_{\alpha}^{Y}$ and $\tau_{\alpha}^{Y'}$.

\section{Conclusions}
\label{sec:4}

The recently developed mode-coupling theory (MCT) for molecular systems
has been applied to calculate the standard correlation functions
which demonstrate the evolution of glassy dynamics of 
a symmetric-hard-dumbbell system (HDS).
The equilibrium structure of this system is determined solely 
by excluded-volume effects, i.e., the sluggish dynamics
and the structural arrest are due to steric hindrance for
translational and reorientational motions.
The phase diagram demonstrates that there are two scenarios for
the liquid-glass-transition dynamics:
one deals with strong and the other with weak steric
hindrance for reorientational motion~\cite{Chong-MCT-dumbbell-1}.

For the strong-steric-hindrance scenario, as it is obtained
for dumbbells of elongation $\zeta$ exceeding, say, 0.6,
the constituent atoms are localized in cages of a similar size as found
for the simple hard-sphere system (HSS) and for the motion of a single dumbbell 
in the HSS, Fig.~\ref{fig:rA-rC-vs-zeta}.
The parameter $\lambda$, which determines the anomalous
exponents of the decay laws and the critical time scales,
is close to that of the HSS, Fig.~\ref{fig:lambda}.
The range of validity of the scaling laws for the
$\beta$- and $\alpha$-processes,
exemplified in the lower panels of 
Figs.~\ref{fig:C1s-beta} and \ref{fig:C1s-alpha}, respectively,
is similar to what was found for the
HSS~\cite{Franosch97,Fuchs98} and for the motion of 
a hard dumbbell in the HSS~\cite{Chong01,Chong01b}.
The rotation-translation coupling implies 
for the mean-squared displacement of the constituent atom
a crossover interval
between the end of the von Schweidler process 
and the beginning of the diffusion process
which is about an order of magnitude larger than for 
the motion of a sphere in the HSS. 
This effect is even more pronounced for the dumbbell liquid,
the lower panel of Fig.~\ref{fig:MSD-scaling},
than for a dumbbell moving in the HSS~\cite{Chong01b}.
Testing this prediction by a molecular-dynamics simulation would 
provide a valuable information on the relevance of our theory.

The results of this paper together with the preceding findings
on the $\alpha$-peaks for reorientational motion for 
angular-momentum index $\ell = 1$ and $\ell = 2$
as well as for the elastic modulus~\cite{Chong01}
lead to the conclusion that the strong-steric-hindrance
scenario explains the qualitative features of the 
structural relaxation in glass-forming van-der-Waals systems like
orthoterphenyl, Salol, or propylene carbonate.
This holds with two reservations.
First, the calculated wave-vector dependence of the Debye-Waller
factor $f_{q}^{N}$ is stronger than that measured by 
neutron-scattering spectroscopy for 
orthoterphenyl~\cite{Toelle98}.
It remains to be shown that application of the theory to 
molecules more complicated than dumbbells will
lead to a smearing out of the strong $q$-variations
characteristic of hard-sphere-like systems.
Second, molecules which allow for a measurement of the 
dipole-correlator by dielectric loss spectroscopy carry
an electric dipole moment.
This leads to long-ranged interactions between the molecules. 
It remains to be studied how the incorporation of these 
interactions change the results based solely on hard-core interactions.

The dipole-susceptibility spectra for a dumbbell with
large $\zeta$ moving in a HSS~\cite{Chong01} obey the scaling proposed
by Dixon {\em et al.}~\cite{Dixon90}.
We found this is the case also for the spectra of the HDS for
elongations in the range $0.6 < \zeta \lesssim 0.8$. 
For $\zeta = 1.0$, on the other hand, the spectra for the high-frequency wing are
below the master curve found in Ref.~\onlinecite{Dixon90},
and for $\zeta = 0.6$ they are slightly above.
Thus, our theory is not consistent with the conjecture that
the quoted scaling law holds universally.  

Decreasing the elongation, the steric-hindrance effects for
reorientations weaken.
A new scenario emerges for sufficiently small $\zeta$ for two
reasons.
First, there is a critical elongation $\zeta_{c}$ separating
a normal glass from a plastic one.
For $\epsilon_{A} = (\zeta - \zeta_{c}) / \zeta_{c}$ approaching
zero, the critical amplitudes $h_{q,s}^{Z}$, in particular
$h_{1,s} = \lim_{q \to 0} h_{q,s}^{Z}$, diverge.
This transition is due to the top-down symmetry of the molecules.
But the asymptotic formulas for small $\epsilon_{A}$ remain
valid also for nearly symmetric molecules;
one merely has to replace $|\, \epsilon_{A} \,|$ by
$\sqrt{ \epsilon_{A}^{2} + \delta^{2} }$, where
$\delta$ is a number quantifying the symmetry 
breaking~\cite{Franosch94,Franosch98b}.
The precursor phenomena of the approach of $\zeta$ towards $\zeta_{c}$
have been discussed before for the motion of a single dumbbell
in a HSS~\cite{Chong01}. 
The most obvious one is the speeding up of the
$\alpha$-process for reorientations relative to that for 
density fluctuations, as demonstrated in 
Fig.~\ref{fig:NZ-t-weak}.
The second reason is the strong increase of the exponent parameter
$\lambda$ for $\zeta \sim \zeta_{c}$, Fig.~\ref{fig:lambda}.
For $\lambda = 1$, there occurs a higher-order glass transition
singularity, where the asymptotic expansions of the MCT solutions
are utterly different from those cited in Sec.~\ref{subsec:2B}.
The approach of $\lambda$ towards unity leads to a shrinking
of the range of validity of the discussed universal formulas.
The corrections diverge, as was discussed in connection with
the coefficient $B_{1}$ in Eq.~(\ref{eq:C1s-alpha-correction}).
This equation shows that the increase of the product $B_{1} h_{1,s}$
leads to an increasing violation of the superposition principle
for the $\alpha$-process, as shown in the upper panel of
Fig.~\ref{fig:C1s-alpha}.
As a result, the $\alpha$-relaxation time approaches the universal
law, Eq.~(\ref{eq:scale-coupling}),
only for very small distance parameters
$\epsilon = (\varphi - \varphi_{c}(\zeta)) / \varphi_{c}(\zeta)$.
For $\zeta = 1.0$, the power-law regime is reached for
$|\, \epsilon \,| = 10^{-1}$, while $|\, \epsilon \,|$ has to be
smaller than $10^{-2}$ for the approach to the
asymptotic law for $\zeta = 0.4$, Fig.~\ref{fig:scaling-time}.
A decrease of $|\, \epsilon \,|$ by an order of magnitude is equivalent
to an increase of the relaxation time by more than a factor of 1000.

For $\zeta = 0.4$, the variation of the $\alpha$-relaxation-time
scales over more than three orders of magnitude can be described well by
power laws, Fig.~\ref{fig:scaling-time-effective}.
But, the above-explained corrections to the leading-order asymptotic
laws imply that the fitted effective exponents $\gamma^{\rm eff}$
are considerably smaller than the exponent $\gamma = 3.77$
specifying the correct asymptotic behavior.
Moreover, the corrections depend on the variable
considered and therefore the found effective exponents are
pairwise different.
This means that the $\alpha$-relaxation scales are not coupled,
rather the ratio of two scales varies according to a power law
as shown for two cases in Fig.~\ref{fig:D-tau}.
Molecular-dynamics-simulation studies for a binary
Lennard-Jones system which have been started by 
Kob and Andersen~\cite{Kob94} have been used for detailed tests
of MCT, as can be inferred from Ref.~\onlinecite{Kob99}
and the papers cited there.
The range of time variations for the $\alpha$-process in these
studies is about the one considered in
Fig.~\ref{fig:scaling-time-effective}.
As a remarkable deviation from the $\alpha$-scale universality,
a deviation of the fit exponent for the diffusivity
from the exponent derived from density-relaxation curves has been
reported~\cite{Kob94}.
Indeed, the $D \tau_{\alpha}$-diagram for the simulation data
is in semi-quantitative agreement with the results shown in
Fig.~\ref{fig:D-tau}~\cite{Kob-private}.
Therefore, it is tempting to conjecture that the simulation data
can be explained as done above for the results in 
Figs.~\ref{fig:scaling-time-effective} and \ref{fig:D-tau}.

We are not aware of experiments which exhibit the weak steric hindrance
scenario.
Molecular-dynamics simulations for a liquid of symmetric Lennard-Jones
dumbbells with elongation $\zeta = 0.33$ lead to the 
suggestion~\cite{Ma97} that an inclusion of the rotational degrees
of freedom is decisive for an understanding of the exponent parameter.
This conclusion is corroborated by Fig.~\ref{fig:lambda}.
Detailed simulation studies of the glassy dynamics of a liquid
of slightly asymmetric Lennard-Jones dumbbells for $\zeta = 0.5$
have been reported by 
K\"ammerer {\em et al.}~\cite{Kaemmerer97,Kaemmerer98,Kaemmerer98b}.
The correlation functions dealing with the translational degrees of
freedom for large and intermediate wave vectors 
and also the ones for reorientational fluctuations for
large angular-momentum indices
could be interpreted with the universal MCT formulas.
However, the dipole correlator did not show a two-step-relaxation
scenario and it exhibited strong violations of the 
superposition principle quite similar to what is shown in 
the upper panel of Fig.~\ref{fig:C1s-alpha}.
The $\alpha$-relaxation time scales for density fluctuations of 
intermediate wave vectors, for the diffusivity, and for the
dipole correlator could be fitted well by power laws with the
exponents 2.56, 2.20, and 1.66, respectively.
The differences in these exponents are quite similar to what is
demonstrated in Fig.~\ref{fig:scaling-time-effective}
for the corresponding quantities.
This explains why the $D \tau_{\alpha}$--versus--temperature
diagram for the simulation results~\cite{Latz-private}
shows violations of the scale coupling in semi-quantitative 
agreement with the ones shown in Fig.~\ref{fig:D-tau}.
It seems that the simulation results for $C_{1,s}(t)$
also fit nicely into the framework of the ideal MCT
for molecular liquids. 

Let us notice that the density correlators of the glass states 
for wave vector $q=9.8$ exhibit oscillations for times around 0.1, 
Figs.~\ref{fig:NZ-t-weak} and \ref{fig:NZ-t-strong}. 
These are the analogues of the oscillations analyzed previously 
for the HSS in connection with a discussion of the so-called 
boson-peak phenomenon and high-frequency sound~\cite{Goetze00}.
It should be mentioned that the dynamics of a dipolar-hard-sphere 
system was analyzed recently within the 
mode-coupling theory~\cite{Theenhaus01} describing the structure 
by tensor-density fluctuations. 
Some comments on the general relation between this theory and the one 
used in the present paper can be found in 
Refs.~\onlinecite{Chong-MCT-dumbbell-1} and \onlinecite{Theenhaus01}.
For the dipolar-hard-sphere system, the oscillations have been analyzed 
in detail. 
They reflect subtle couplings between translational and rotational 
degrees of freedom as enforced by the long-ranged Coulomb interactions. 
For the HDS such couplings are present as well, albeit caused by the 
short-ranged steric hindrance effects; 
but an analysis of these oscillations remains to be done. 

\begin{acknowledgments}

We thank W.~Kob cordially for discussions and suggestions.
We thank him as well as M.~Sperl, Th.~Voigtmann, and R.~Schilling
for helpful critique of the manuscript.

\end{acknowledgments}

\end{document}